\documentclass[journal]{IEEEtran}

\ifCLASSINFOpdf
\else
\fi
\usepackage{amsbsy}
\usepackage{amsmath}
\usepackage{fixmath}
\usepackage{cancel}
\usepackage{amssymb}
\usepackage{epsfig}
\usepackage{epstopdf}
\usepackage{algorithmic}
\usepackage{algorithm}

\usepackage{tikz}
\usetikzlibrary{arrows,decorations,backgrounds,shadows,plotmarks, positioning, calc}

\usepackage{pgfplots}
\pgfplotsset{compat=newest}
\pgfplotsset{plot coordinates/math parser=false}
\pgfplotsset{every axis/.append style={font=\footnotesize}}
\pgfplotsset{
    ylabel right/.style={
        after end axis/.append code={
            \node [rotate=90, anchor=north] at (rel axis cs:1,0.5) {#1};
        }   
    }
}
\newlength\figureheight
\newlength\figurewidth
\newlength\subgraphheight
\newlength\subgraphwidth
\makeatletter
\newcommand{\nodedistance}{\tikz@node@distance}
\makeatother

\makeatletter

\pgfkeys{
  /tikz/blockm/.initial=1,
  /tikz/blockn/.initial=1
}

\pgfdeclareshape{dspantenna}{

	\inheritsavedanchors[from=rectangle]

    \inheritanchorborder[from=rectangle]

    \inheritanchor[from=rectangle]{center}
    \inheritanchor[from=rectangle]{south}
    \inheritanchor[from=rectangle]{west}
    \inheritanchor[from=rectangle]{north}
    \inheritanchor[from=rectangle]{east}
    \inheritanchor[from=rectangle]{south east}
    \inheritanchor[from=rectangle]{south west}
    \inheritanchor[from=rectangle]{north east}
    \inheritanchor[from=rectangle]{north west}
	\saveddimen{\halfheight}{
	\pgfmathsetlength{\pgf@xa}{\pgfkeysvalueof{/pgf/minimum height}/2}
	\pgfmathsetlength{\pgf@xb}{\pgfkeysvalueof{/pgf/outer ysep}}
	\advance\pgf@xa by \pgf@xb
	\pgf@x=\pgf@xa
	}
	\saveddimen{\halfwidth}{
		\pgfmathsetlength{\pgf@xa}{\pgfkeysvalueof{/pgf/minimum width}/2}
		\pgfmathsetlength{\pgf@xb}{\pgfkeysvalueof{/pgf/outer xsep}}
		\advance\pgf@xa by \pgf@xb
		\pgf@x=\pgf@xa
		}
		
    \backgroundpath{

        \northeast \pgf@xa=\pgf@x \pgf@ya=\pgf@y
                \pgf@xb=\pgf@xa \pgf@yb=\pgf@ya
        
        \pgfpathmoveto{\pgfpoint{\pgf@xb}{\pgf@yb}}
        \advance \pgf@xb by -\halfwidth
        \advance \pgf@yb by -\halfwidth
        \pgfpathlineto{\pgfpoint{\pgf@xb}{\pgf@yb}}
        \advance \pgf@xb by -\halfwidth
        \advance \pgf@yb by \halfwidth
        \pgfpathlineto{\pgfpoint{\pgf@xb}{\pgf@yb}}
        \advance \pgf@xb by \halfwidth
        \advance \pgf@yb by -\halfwidth
        \pgfpathmoveto{\pgfpoint{\pgf@xb}{\pgf@yb}}
        \advance \pgf@yb by -\halfwidth
        \pgfpathlineto{\pgfpoint{\pgf@xb}{\pgf@yb}}
        \advance \pgf@xb by \halfwidth
        \pgfpathlineto{\pgfpoint{\pgf@xb}{\pgf@yb}}
        
        \pgfusepath{draw}
    }
}

\pgfdeclareshape{dspvdots}{

	\inheritsavedanchors[from=rectangle]

    \inheritanchorborder[from=rectangle]

    \inheritanchor[from=rectangle]{center}
    \inheritanchor[from=rectangle]{south}
    \inheritanchor[from=rectangle]{west}
    \inheritanchor[from=rectangle]{north}
    \inheritanchor[from=rectangle]{east}
    \inheritanchor[from=rectangle]{south east}
    \inheritanchor[from=rectangle]{south west}
    \inheritanchor[from=rectangle]{north east}
    \inheritanchor[from=rectangle]{north west}
	\saveddimen{\halfheight}{
	\pgfmathsetlength{\pgf@xa}{\pgfkeysvalueof{/pgf/minimum height}/2}
	\pgfmathsetlength{\pgf@xb}{\pgfkeysvalueof{/pgf/outer ysep}}
	\advance\pgf@xa by \pgf@xb
	\pgf@x=\pgf@xa
	}
	\saveddimen{\halfwidth}{
		\pgfmathsetlength{\pgf@xa}{\pgfkeysvalueof{/pgf/minimum width}/2}
		\pgfmathsetlength{\pgf@xb}{\pgfkeysvalueof{/pgf/outer xsep}}
		\advance\pgf@xa by \pgf@xb
		\pgf@x=\pgf@xa
		}
		
    \backgroundpath{

        \northeast \pgfutil@tempdima=\pgf@x \pgfutil@tempdimb=\pgf@y
		
		\pgf@xa=\halfwidth
		\pgf@ya=\halfheight
		\advance\pgfutil@tempdima by -\pgf@xa
		\advance\pgfutil@tempdimb by -\pgf@ya 
		
       \pgfpathcircle{\pgfpoint{\pgfutil@tempdima}{\pgfutil@tempdimb}}{\halfwidth/3}
        \pgf@xa=\halfwidth
        \pgf@ya=\halfheight
       \advance\pgfutil@tempdimb by \pgf@xa
       \advance\pgfutil@tempdimb by -\pgf@ya
        \pgfpathcircle{\pgfpoint{\pgfutil@tempdima}{\pgfutil@tempdimb}}{\halfwidth/3}
        \pgf@xa=\halfwidth
        \pgf@ya=\halfheight
        \advance\pgfutil@tempdimb by 2\pgf@ya
        \advance\pgfutil@tempdimb by -2\pgf@xa
        \pgfpathcircle{\pgfpoint{\pgfutil@tempdima}{\pgfutil@tempdimb}}{\halfwidth/3} 
        
        \pgfusepath{fill}
    }
}

\pgfdeclareshape{dspadder}{
\inheritsavedanchors[from=circle]
    \inheritanchorborder[from=circle]

    \inheritanchor[from=circle]{center}
    \inheritanchor[from=circle]{south}
    \inheritanchor[from=circle]{west}
    \inheritanchor[from=circle]{north}
    \inheritanchor[from=circle]{east}
    \inheritanchor[from=circle]{south east}
    \inheritanchor[from=circle]{south west}
    \inheritanchor[from=circle]{north east}
    \inheritanchor[from=circle]{north west}

    \inheritbackgroundpath[from=circle]

    \beforebackgroundpath{

                \radius \pgf@xb=\pgf@x
        \centerpoint \pgf@xa=\pgf@x \pgf@ya=\pgf@y \pgf@xc=\pgf@x \pgf@yc=\pgf@y

        \advance\pgf@yc by \pgf@xb
        \pgfpathmoveto{\pgfpoint{\pgf@xc}{\pgf@yc}}
        \advance\pgf@yc by -2\pgf@xb
        \pgfpathlineto{\pgfpoint{\pgf@xc}{\pgf@yc}}
        \advance\pgf@xc by -\pgf@xb
        \advance\pgf@yc by \pgf@xb
        \pgfpathmoveto{\pgfpoint{\pgf@xc}{\pgf@yc}}
        \advance\pgf@xc by 2\pgf@xb
        \pgfpathlineto{\pgfpoint{\pgf@xc}{\pgf@yc}}

        \pgfusepath{draw}
    }
}

\pgfdeclareshape{dsptriangler}{

	\inheritsavedanchors[from=rectangle]

    \inheritanchorborder[from=rectangle]

    \inheritanchor[from=rectangle]{center}
    \inheritanchor[from=rectangle]{south}
    \inheritanchor[from=rectangle]{west}
    \inheritanchor[from=rectangle]{north}
    \inheritanchor[from=rectangle]{east}
    \inheritanchor[from=rectangle]{south east}
    \inheritanchor[from=rectangle]{south west}
    \inheritanchor[from=rectangle]{north east}
    \inheritanchor[from=rectangle]{north west}
	\saveddimen{\halfheight}{
	\pgfmathsetlength{\pgf@xa}{\pgfkeysvalueof{/pgf/minimum height}/2}
	\pgfmathsetlength{\pgf@xb}{\pgfkeysvalueof{/pgf/outer ysep}}
	\advance\pgf@xa by \pgf@xb
	\pgf@x=\pgf@xa
	}
	\saveddimen{\halfwidth}{
		\pgfmathsetlength{\pgf@xa}{\pgfkeysvalueof{/pgf/minimum width}/2}
		\pgfmathsetlength{\pgf@xb}{\pgfkeysvalueof{/pgf/outer xsep}}
		\advance\pgf@xa by \pgf@xb
		\pgf@x=\pgf@xa
		}
		
    \backgroundpath{

        \pgfutil@tempdima=\halfheight
        \pgfutil@tempdimb=\halfwidth
        \southwest \pgf@xa=\pgf@x \pgf@ya=\pgf@y 
		\pgfpathmoveto{\pgfpoint{\pgf@xa}{\pgf@ya}}
		\advance \pgf@ya by 2\pgfutil@tempdima
     	\pgfpathlineto{\pgfpoint{\pgf@xa}{\pgf@ya}}
     	\advance \pgf@xa by 2\pgfutil@tempdimb
     	\advance \pgf@ya by -\pgfutil@tempdima
     	\pgfpathlineto{\pgfpoint{\pgf@xa}{\pgf@ya}}	
     	\advance \pgf@xa by -2\pgfutil@tempdimb
     	\advance \pgf@ya by -\pgfutil@tempdima
     	\pgfpathlineto{\pgfpoint{\pgf@xa}{\pgf@ya}}       
        \pgfusepath{draw}
    }
}

\pgfdeclareshape{dspdiamond}{

	\inheritsavedanchors[from=rectangle]

    \inheritanchorborder[from=rectangle]

    \inheritanchor[from=rectangle]{center}
    \inheritanchor[from=rectangle]{south}
    \inheritanchor[from=rectangle]{west}
    \inheritanchor[from=rectangle]{north}
    \inheritanchor[from=rectangle]{east}
    \inheritanchor[from=rectangle]{south east}
    \inheritanchor[from=rectangle]{south west}
    \inheritanchor[from=rectangle]{north east}
    \inheritanchor[from=rectangle]{north west}
	\saveddimen{\halfheight}{
	\pgfmathsetlength{\pgf@xa}{\pgfkeysvalueof{/pgf/minimum height}/2}
	\pgfmathsetlength{\pgf@xb}{\pgfkeysvalueof{/pgf/outer ysep}}
	\advance\pgf@xa by \pgf@xb
	\pgf@x=\pgf@xa
	}
	\saveddimen{\halfwidth}{
		\pgfmathsetlength{\pgf@xa}{\pgfkeysvalueof{/pgf/minimum width}/2}
		\pgfmathsetlength{\pgf@xb}{\pgfkeysvalueof{/pgf/outer xsep}}
		\advance\pgf@xa by \pgf@xb
		\pgf@x=\pgf@xa
		}
		
    \backgroundpath{

        \pgfutil@tempdima=\halfheight
        \pgfutil@tempdimb=\halfwidth
        \southwest \pgf@xa=\pgf@x \pgf@ya=\pgf@y 
        \advance\pgf@xa by \pgfutil@tempdimb
		\pgfpathmoveto{\pgfpoint{\pgf@xa}{\pgf@ya}}
		\advance \pgf@xa by \pgfutil@tempdimb
		\advance \pgf@ya by \pgfutil@tempdima
     	\pgfpathlineto{\pgfpoint{\pgf@xa}{\pgf@ya}}
     	\advance \pgf@xa by -\pgfutil@tempdimb
     	\advance \pgf@ya by \pgfutil@tempdima
     	\pgfpathlineto{\pgfpoint{\pgf@xa}{\pgf@ya}}	     	
     	\advance \pgf@xa by -\pgfutil@tempdimb
     	\advance \pgf@ya by -\pgfutil@tempdima
     	\pgfpathlineto{\pgfpoint{\pgf@xa}{\pgf@ya}}      	
     	\advance \pgf@xa by \pgfutil@tempdimb
     	\advance \pgf@ya by -\pgfutil@tempdima
     	\pgfpathlineto{\pgfpoint{\pgf@xa}{\pgf@ya}}     
        \pgfusepath{draw}
    }
}

\makeatother

\usepackage{pgfplots}
\usepackage{pgfkeys}
\pgfplotsset{compat=newest}
\pgfplotsset{plot coordinates/math parser=false}

\DeclareMathOperator{\sign}{sign}

\DeclareMathOperator{\dB}{dB}

\DeclareMathAlphabet{\mathbit}{OML}{cmr}{bx}{it}
\DeclareMathAlphabet{\mathsf}{OT1}{cmss}{m}{n}
\DeclareMathAlphabet{\mathbsf}{OT1}{cmss}{bx}{it}
\newcommand{\btm}[1]{\boldsymbol{\tilde{\mathcal{#1}}}} 
\newcommand{\bhm}[1]{\boldsymbol{\hat{\mathcal{#1}}}} 
\newcommand{\bm}[1]{\boldsymbol{\mathcal{#1}}} 
\newcommand{\inC}[1]{\ensuremath{\in\mathbb{C}^{#1}}}
\newcommand{\inR}[1]{\ensuremath{\in\mathbb{R}^{#1}}}
\newcommand{\inset}[2]{\ensuremath{\in \left\{#1,\ldots,#2\right\}}}
\newcommand{\Real}[1]{\ensuremath{\Re\left\{#1\right\}}}
\newcommand{\Imag}[1]{\ensuremath{\Im\left\{#1\right\}}}
\newcommand\given[1][]{\:#1\vert\:}

\newcommand\diff[1]{\ensuremath{\:\mathrm{d}#1}}

\begin{document}
%
\title{Channel Estimation and Data Equalization in Frequency-Selective MIMO Systems with One-Bit Quantization}
%
%
%

\author{Javier Garc\'ia,~\IEEEmembership{Student Member,~IEEE,}
        Jawad Munir,~Kilian Roth~
        and~Josef A. Nossek,~\IEEEmembership{Life~Fellow,~IEEE}
}

\maketitle

\begin{abstract}
This paper addresses channel estimation and data equalization on frequency-selective 1-bit quantized Multiple Input-Multiple Output (MIMO) systems. No joint processing or Channel State Information is assumed at the transmitter, and therefore our findings are also applicable to the uplink of Multi-User MIMO systems. System models for both Orthogonal Division Frequency Multiplexing (OFDM) and single-carrier schemes are developed. A Cram\'er-Rao Lower Bound for the estimation problems is derived. The two nonlinear algorithms Expectation Maximization (EM) and Generalized Approximate Message Passing (GAMP) are adapted to the problems, and a linear method based on the Bussgang theorem is proposed. In the OFDM case, the linear method enables subcarrier-wise estimation, greatly reducing computational complexity. Simulations are carried out to compare the algorithms with different settings. The results turn out to be close to the Cram\'er-Rao bound in the low Signal to Noise Ratio (SNR) region. The OFDM setting is more suitable for the nonlinear algorithms, and that the linear methods incur a performance loss with respect to the nonlinear approaches. In the relevant low and medium SNR regions, the loss amounts to 2-3 dB and might well be justified in exchange for the reduced computational effort, especially in Massive MIMO settings.
\end{abstract}


%
\IEEEpeerreviewmaketitle

\section{Introduction}
%
%
%
%
\IEEEPARstart{T}{he} fifth generation (5G) of mobile communications is expected to increase spectral and energy efficiency by several orders of magnitude~\cite{wang_5g}. To fulfill this requirement, among other technologies, Multiple Input-Multiple Output (MIMO) systems with large numbers of antennas ~\cite{larsson_massve_mimo} are being considered. They which would considerably increase array gain. Higher frequency bands~\cite{boccardi_5g_hf}, (6-100 GHz) are also being investigated. These would allow for larger bandwidth.

These changes place stringent requirements on the receiver-side analog-to-digital converters (ADCs). Due to the high frequency and large bandwidth, the ADCs need to operate at high sampling rate, which leads to high power consumption. This problem is increased in a Massive MIMO setting, which requires a large amount of ADCs.

The power consumption of the ADCs grows exponentially with the number of bits, as shown in~\cite{murmann2015adc} and~\cite{walden_adc}. Therefore, the low-resolution (1 to 3 bits) ADCs have been proposed as a way to address the power consumption problem. We focus on the 1-bit quantization case.

Various aspects of the 1-bit quantized MIMO channel have been analyzed in recent work. An analysis in~\cite{mo_mimo_1b_capacity} and~\cite{roth_mimo_1b_capacity} shows that, if the number of antennas of the $1$-bit quantized system is increased to the point in which it consumes the same power as the unquantized one, the capacity of the quantized system can beat that of the unquantized one at the low and medium SNR regions.

Different channel estimation algorithms are discussed in~\cite{mo_mimo_1b_chest}. They obtain good results but are iterative and nonlinear. Furthermore, convergence is not guaranteed, especially if the channel taps are not i.i.d. Gaussian, which is the case in practical scenarios.

A linear MMSE receiver for equalization of quantized MIMO channels is proposed in~\cite{mezghani_mmse_receiver}, and an iterative nonlinear one in~\cite{mezghani_mimo_1b_dfe}. The nonlinear equalizer achieves better BER performance at the cost of increased computational complexity.

Finally, the problem of joint channel and data estimation (JCD) is treated in~\cite{wen_jcd}. The authors show that this approach greatly improves the results, and requires fewer pilots.

One of the main shortcomings of the mentioned contributions is that they consider only flat-fading channels. With unquantized systems, this assumption can be justified by the use of multi-carrier modulations, such as Orthogonal Frequency Division Multiplexing (OFDM), because then the channel in each subcarrier is flat. However, in the quantized case, the subcarriers can no longer be separated without loss. We note that OFDM is still attractive in this case because it allows for uplink multiplexing with optimal channel allocation.

To the best of our knowledge, very little work has been done on quantized frequency-selective MIMO channels or on the loss incurred by using multi-carrier modulations. A model for channel estimation and equalization in quantized MIMO OFDM systems is proposed in~\cite{studer_mimo_1b_ofdm}. However, it relies on convex optimization algorithms, which, for the large-dimensional problems at hand, have high computational cost.

In this paper, we develop a model for channel estimation and for equalization of both single-carrier and OFDM quantized MIMO systems. We derive a Cram\'er-Rao lower bound for the estimation problem, that can be used as a benchmark for algorithm performance comparison. We adapt the existing nonlinear iterative algorithms Expectation Maximization (EM) and Generalized Approximate Message Passing (GAMP) to solve these estimation problems in the minimum mean square error (MMSE) sense. Additionally, we develop a linear estimator based on the Bussgang theorem, which greatly reduces the computational complexity in the OFDM case because it allows for per-subcarrier equalization. Through simulations, we then compare the performance of all the estimation methods in different scenarios, and draw some important conclusions. 

We note that all our analysis does not assume any joint processing or Channel State Information (CSI) at the transmitter. Therefore, our findings are also applicable to Multi-User MIMO uplink channels, by considering each transmit antenna (or group of them) as a separate user.

The paper is organized as follows. In Section~\ref{sec:model}, the estimation problem formulation for both OFDM and single-carrier quantized MIMO is developed. Section~\ref{sec:theory} derives the Cram\'er-Rao Lower Bound for this problem and presents the nonlinear and linear algorithms to solve it. This section also analyzes the computational complexity of the methods. In Section~\ref{sec:results}, the algorithms are compared with the use of simulations. Finally, Section~\ref{sec:conclusions} summarizes the most important results and identifies some areas for future work.

\section{System Model}\label{sec:model}
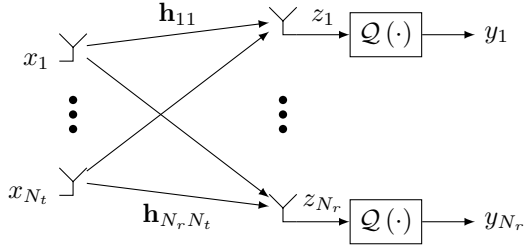
\begin{figure}[!t]\centering
\begin{tikzpicture}[
 node distance=2em,
 antenna/.style={
 	dspantenna,
 	minimum width=1em,
 	minimum height=1em,
 	draw,
 },
 vdots/.style={
    dspvdots,
    minimum height=2em,
    minimum width=0.9em,
    draw
 },
 block/.style={
 	rectangle,
 	minimum width=2.5em,
 	minimum height=1.55em,
 	draw
 }
]
\node (txdots) [vdots] {};
\node (t1) [antenna, xscale=-1, above=0.5*\nodedistance of txdots] {};
\node (tN) [antenna, xscale=-1, below=0.5*\nodedistance of txdots] {};

\node at (t1.south east) [anchor=east] {$x_1$};
\node at (tN.south east) [anchor=east] {$x_{N_t}$};

\node (rxdots) [vdots, right=3.5*\nodedistance of txdots] {};
\node (r1) [antenna, above=of rxdots] {};
\node (rN) [antenna, below=of rxdots] {};
\node (Q1) [block, right=of r1.south east] {$\mathcal{Q}\left(\cdot\right)$};
\node (QN) [block, right=of rN.south east] {$\mathcal{Q}\left(\cdot\right)$};

\draw[arrows={-latex}] (r1.south east) -- node[above] {$z_1$} (Q1.west);
\draw[arrows={-latex}] (rN.south east) -- node[above] {$z_{N_r}$} (QN.west);
\draw[arrows={-latex}] (Q1.east) -- ++ (\nodedistance, 0) node[right] {$y_1$};
\draw[arrows={-latex}] (QN.east) -- ++ (\nodedistance, 0) node[right] {$y_{N_r}$};

\draw [arrows={-latex}] (t1) -- node [above] {$\mathbf{h}_{11}$} (r1);
\draw [arrows={-latex}] (t1) -- (rN);
\draw [arrows={-latex}] (tN) -- (r1);
\draw [arrows={-latex}] (tN) -- node [below] {$\mathbf{h}_{N_rN_t}$} (rN);
\end{tikzpicture}
\caption{Quantized frequency-selective MIMO channel} \label{fig:qmimo}
\end{figure}
\subsection{OFDM System Model}
We consider a MIMO system with $N_r$ receive antennas, $N_t$ transmit antennas, and receiver-side 1-bit quantization (Fig.~\ref{fig:qmimo}).

First, we will develop a system model for the OFDM~\cite{van_Nee_OFDM_book} case, with $N$ subcarriers. No CSI is assumed at the transmitter. At each transmit antenna $n_t\inset{1}{N_t}$, a different sequence of $M$ OFDM symbols, $\mathbf{X}_{n_t}\inC{N\times M}$ is generated.

We consider an arbitrary frequency-selective MIMO channel. Each pair of transmit antenna $n_t\inset{1}{N_t}$ and receive antenna $n_r\inset{1}{N_r}$ has a channel impulse response of $L$ taps, denoted by $\mathbf{h}_{n_rn_t}\inC{L}$.

Due to the use of a cyclic prefix (CP), each pair of antennas $\left(n_r,n_t\right)\in\left\{1,\ldots N_r\right\}\times\left\{1,\ldots,N_t\right\}$ has an equivalent circulant channel convolution matrix $\mathbf{H}_{n_rn_t}\inC{N\times N}$ whose first column is $\left[\mathbf{h}_{n_rn_t}^T \mathbf{0}_{(N-L)\times 1}^T\right]^T$.

After the application of an IFFT and the channel, the unquantized received signals $\mathbf{Z}_{n_r}\inC{N\times M}$ (where $n_r\inset{1}{N_r}$ is the receive antenna) are:
\begin{equation}
\mathbf{Z}_{n_r}=\sum_{n_t=1}^{N_t}\mathbf{H}_{n_rn_t}\mathbf{F}^H\mathbf{X}_{n_t}+\mathbf{W}_{n_r}.
\end{equation}
Here, $\mathbf{F}$ denotes a \textit{unitary} $N\times N$ DFT matrix. The noise $\mathbf{W}_{n_r}\inC{N\times M}$ is additive Gaussian and spatially and temporally uncorrelated. Its samples have variances $\sigma_{n_rnm}^2$.

The receiver then applies 1-bit quantization to $\mathbf{Z}_{n_r}$, and obtains the observations $\mathbf{Y}_{n_r}\inC{N\times M}$:
\begin{equation}
\mathbf{Y}_{n_r}=\mathcal{Q}\left(\sum_{n_t=1}^{N_t}\mathbf{H}_{n_rn_t}\mathbf{F}^H\mathbf{X}_{n_t}+\mathbf{W}_{n_r}\right),
\label{eq:Y_nr_pre}
\end{equation}
where the operator $\mathcal{Q}\left(\cdot\right)$ is applied element-wise, and takes the sign of the real and imaginary parts of the argument:
\begin{equation}
\mathcal{Q}\left(x\right)\triangleq\sign\left\{\Re\left\{x\right\}\right\}+j\sign\left\{\Im\left\{x\right\}\right\}.
\label{eq:Q}
\end{equation}

The circulant channel convolution matrices $\mathbf{H}_{n_rn_t}$ are diagonalized by DFT matrices, and the resulting diagonal matrix is the $N$-point DFT of the channel impulse response:
\begin{equation}
\mathbf{F}\mathbf{H}_{n_rn_t}\mathbf{F}^H=\mathbf{\Lambda}_{n_rn_t}=\mathrm{diag}\left\{\mathbf{F}_{N\times L}\mathbf{h}_{n_rn_t}\right\},
\end{equation}
where $\mathbf{F}_{N\times L}$ contains the $L$ first columns of an $N$-point DFT matrix \textit{with unit-magnitude entries}. This lets us rewrite (\ref{eq:Y_nr_pre}) as:
\begin{equation}
\mathbf{Y}_{n_r}=\mathcal{Q}\left(\sum_{n_t=1}^{N_t}\mathbf{F}^H\mathbf{\Lambda}_{n_rn_t}\mathbf{X}_{n_t}+\mathbf{W}_{n_r}\right),
\label{eq:model}
\end{equation}.

\subsubsection{Problem Formulation for OFDM Channel Estimation}
With orthogonal pilots, the channel estimation problem is independent across receive antennas. Consider a sequence of $T$ pilot blocks $\mathbf{X}_{n_t}\inC{N\times T}$, with $n_t\inset{1}{N_t}$. Vectorizing the signal $\mathbf{Y}_{n_r}$ at each receive antenna in (\ref{eq:model}) gives:
\begin{equation}
\mathbf{y}_{n_r}=\mathcal{Q}\left(\sum_{n_t=1}^{N_t}\left(\mathbf{X}_{n_t}^T\diamond\mathbf{F}^H\right)\mathbf{F}_{N\times L}\mathbf{h}_{n_rn_t}+\mathbf{w}_{n_r}\right),
\label{eq:y_nr_vec}
\end{equation}
where $\mathbf{y}_{n_r}\triangleq\mathrm{vec}\left(\mathbf{Y}_{n_r}\right)$, and $\mathbf{w}_{n_r}\triangleq\mathrm{vec}\left(\mathbf{W}_{n_r}\right)$, and $\diamond$ denotes the Khatri-Rao product (column-wise Kronecker product). Here, we have used the following property of the vectorization operator:
\begin{equation}
\mathrm{vec}\left(\mathbf{B}\mathrm{diag}\left\{\mathbf{c}\right\}\mathbf{D}\right)=\left(\mathbf{D}^T\diamond\mathbf{B}\right)\mathbf{c}.
\end{equation}

We define the vector $\mathbf{h}_{n_r}\inC{LN_t\times 1}$ as:
\begin{equation}
\mathbf{h}_{n_r}\triangleq\left(\begin{array}{c c c c}
\mathbf{h}_{n_r1} \\
\mathbf{h}_{n_r2} \\
\vdots \\
\mathbf{h}_{n_rN_t} \\
\end{array}\right).
\label{eq:model_H}
\end{equation}
Furthermore, we define the matrix $\mathbf{A}\inC{NT\times LN_t}$ as:
\begin{equation}
\mathbf{A}=\left[\begin{array}{ c c c }
\left(\mathbf{X}_{1}^T\diamond\mathbf{F}^H\right)\mathbf{F}_{N\times L}, & \cdots, & \left(\mathbf{X}_{N_t}^T\diamond\mathbf{F}^H\right)\mathbf{F}_{N\times L}
\end{array}
\right].
\label{eq:chest_sensing}
\end{equation}
We can now write (\ref{eq:y_nr_vec}) as:
\begin{equation}
\mathbf{y}_{n_r}=\mathcal{Q}\left(\mathbf{A}\mathbf{h}_{n_r}+\mathbf{w}_{n_r}\right),\quad n_r\inset{1}{N_r},
\label{eq:channel_estimation_model}
\end{equation}
where $\mathbf{w}_{n_r}=\mathrm{vec}\left(\mathbf{W}_{n_r}\right)\inC{NT\times 1}$ contains uncorrelated Gaussian samples with variances $\sigma_{n_tn_r}^2$.

\subsubsection{Problem Formulation for OFDM Data Equalization}
Consider again the model in (\ref{eq:model}). The problem is now independent across the $M$ transmitted symbols. For symbol $m\inset{1}{M}$ we define the vector of unknowns $\mathbf{x}_m\inC{NN_t\times 1}$:
\begin{equation}
\mathbf{x}_m=\left(\begin{array}{c}
\mathbf{X}_{1m\cdot}\\ \vdots \\ \mathbf{X}_{N_tm\cdot}
\end{array}\right),
\label{eq:model_X}
\end{equation}
the sensing matrix $\mathbf{A}\inC{NN_r\times{NN_t}}$:
\begin{equation}
\mathbf{A}=\left(\begin{array}{c c c}
\mathbf{F}^H\mathbf{\Lambda}_{11} & \cdots & \mathbf{F}^H\mathbf{\Lambda}_{1N_t} \\
\vdots & \ddots & \vdots \\
\mathbf{F}^H\mathbf{\Lambda}_{N_r1} & \cdots & \mathbf{F}^H\mathbf{\Lambda}_{N_rN_t}
\end{array}\right),
\label{eq:eq_sensing}
\end{equation}
and the observation vector $\mathbf{y}_m\inC{NN_r\times 1}$:
\begin{equation}
\mathbf{y}_m=\left(\begin{array}{c}
\mathbf{Y}_{1m\cdot}\\ \vdots \\ \mathbf{Y}_{N_rm\cdot}
\end{array}\right).
\label{eq:model_Y_eq}
\end{equation}
Then the model for data equalization is given by:
\begin{equation}
\mathbf{y}_m=\mathcal{Q}\left(\mathbf{A}\mathbf{x}_m+\mathbf{w}_m\right),
\label{eq:data_equalization_model}
\end{equation}
where $\mathbf{w}_m$ is defined in the same way as $\mathbf{y}_m$, and contains uncorrelated complex Gaussian samples with variance $\sigma_{n_rnm}^2$.

\subsection{Single-Carrier System Model}
To enable block processing, a cyclic prefix is also added in the single-carrier (SC) case. The block size is $N$, and the cyclic prefix has length $L$. The received signal at antenna $n_r\inset{1}{N_r}$ can be written as:
\begin{equation}
\mathbf{Y}_{n_r}=\mathcal{Q}\left(\sum_{n_t=1}^{N_t}\mathbf{H}_{n_rn_t}\mathbf{X}_{n_t}+\mathbf{W}_{n_r}\right),
\label{eq:model_sc_pre}
\end{equation}
where $\mathbf{X}_{n_t}\inC{N\times M}$ horizontally stacks $M$ blocks of transmitted symbols, and $\mathbf{H}_{n_rn_t}\inC{N\times N}$ is defined in the same way as in the OFDM case.
\subsubsection{Problem Formulation for SC Channel Estimation}
For channel estimation, each transmit antenna sends $T$ consecutive orthogonal blocks as pilots. We denote the pilot vector at transmit antenna $n_t\inset{1}{N_t}$ and block $t\inset{1}{T}$ as $\mathbf{x}_{n_tt}\inC{N\times 1}$. We further define the partial circulant convolution matrix $\mathbf{X}_{n_tt}\inC{N\times L}$ in the following way:
\begin{equation}
\left[\mathbf{X}_{n_tt}\right]_{n\ell}\triangleq \mathbf{x}_{n_tt}[n-\ell],
\end{equation}
i.e. the first $L$ columns of a circulant matrix whose first column is $\mathbf{x}_{n_tt}$. With this definition, we can express the channel estimation problem for single-carrier as:
\begin{equation}
\mathbf{y}_{n_r}=\mathcal{Q}\left(\mathbf{A}\mathbf{h}_{n_r}+\mathbf{w}_{n_r}\right),
\label{eq:channel_estimation_model_sc}
\end{equation}
where
\begin{equation}
\mathbf{A}=\left(\begin{array}{c c c}
\mathbf{X}_{11}&\cdots&\mathbf{X}_{N_t1}\\
\vdots&\ddots&\vdots\\
\mathbf{X}_{1T}&\cdots&\mathbf{X}_{N_tT}
\end{array}\right)\inC{NT\times LN_t},
\end{equation}
\begin{equation}
\mathbf{y}_{n_r}=\mathrm{vec}\left(\mathbf{Y}_{n_r}\right),
\end{equation}
and $\mathbf{h}_{n_r}$ is given by (\ref{eq:model_H}).

\subsubsection{Problem Formulation for SC Data Equalization}
From (\ref{eq:model_sc_pre}), we can directly write the model:
\begin{equation}
\mathbf{y}_{m}=\mathcal{Q}\left(\mathbf{A}\mathbf{x}_m+\mathbf{w}_m\right),
\label{eq:data_equalization_model_sc}
\end{equation}
where
\begin{equation}
\mathbf{A}=\left(\begin{array}{c c c}
\mathbf{H}_{11} & \cdots & \mathbf{H}_{1N_t} \\
\vdots & \ddots & \vdots \\
\mathbf{H}_{N_r1} & \cdots & \mathbf{H}_{N_rN_t}
\end{array}\right),
\label{eq:eq_sensing_sc}
\end{equation}
and $\mathbf{x}_m$ and $\mathbf{y}_m$ are given by (\ref{eq:model_X}) and (\ref{eq:model_Y_eq}) respectively.

\section{Theoretical derivations}\label{sec:theory}

\subsection{Cram\'er Rao Bound}\label{sec:cramer}
In this section, we derive the Cram\'er-Rao Lower Bound (CRLB) of the four estimation problems in Section~\ref{sec:model}. This will provide a theoretical limit on the estimation accuracy, which can be used as a benchmark.

The four considered problems (\ref{eq:channel_estimation_model}), (\ref{eq:data_equalization_model}), (\ref{eq:channel_estimation_model_sc}) and (\ref{eq:data_equalization_model_sc}) can be expressed with the following generic model:
\begin{equation}
\mathbf{y}=\mathcal{Q}\left(\mathbf{A}\mathbf{h}+\mathbf{w}\right).
\end{equation}
To compute the CRLB, first we make the problem real-valued:
\begin{equation}
\tilde{\mathbf{y}}=\mathcal{Q}\left(\tilde{\mathbf{A}}\tilde{\mathbf{h}}+\tilde{\mathbf{w}}\right),
\label{eq:cr_real_valued}
\end{equation}
where
\begin{equation}
\tilde{\mathbf{h}}=\left(\begin{array}{c}\Real{\mathbf{h}}\\ \Imag{\mathbf{h}}\end{array}\right);\qquad \tilde{\mathbf{A}}=\left(\begin{array}{c c}
\Real{\mathbf{A}} & -\Imag{\mathbf{A}} \\ \Imag{\mathbf{A}} & \Real{\mathbf{A}} \\
\end{array}\right);
\label{eq:make_real}
\end{equation}
and $\tilde{\mathbf{y}}$ and $\tilde{\mathbf{w}}$ are defined in a similar way as $\tilde{\mathbf{h}}$.

The Cram\'er-Rao bound is then given by:
\begin{equation}
\mathbf{C}_{\hat{\tilde{\mathbf{h}}}\hat{\tilde{\mathbf{h}}}}\succeq\tilde{\mathbf{I}}\left(\tilde{\mathbf{h}}\right)^{-1},
\label{eq:cr_bound}
\end{equation}
where $\mathbf{C}\succeq\mathbf{D}$ indicates that $\mathbf{C}-\mathbf{D}$ is positive semidefinite.
The real-valued Fisher information matrix is computed as:
\begin{equation}
\tilde{\mathbf{I}}\left(\tilde{\mathbf{h}}\right)=\tilde{\mathbf{A}}^T\mathrm{diag}\left\{\frac{1}{\tilde{\sigma}_k^2}\frac{\phi\left(\mu_k\right)^2}{\mathrm{\Phi}\left(\mu_k\right)\left(1-\mathrm{\Phi}\left(\mu_k\right)\right)}\right\}_{k=1}^K\tilde{\mathbf{A}},
\label{eq:cr_fisher_final}
\end{equation}
where
\begin{equation}
\mu_k=\frac{1}{\tilde{\sigma}_k}\sum_{p=1}^P \tilde{a}_{kp}\tilde{h}_p.
 \label{eq:cr_mu_k}
\end{equation}
The derivation of $\tilde{\mathbf{I}}\left(\tilde{\mathbf{h}}\right)$ is given in Appendix~\ref{sec:fisher}. To transform it back to the complex domain, we apply the chain rule to get:
\begin{multline}
\mathbf{I}\left(\boldsymbol{\theta}\right)=\frac{1}{4}\left(\left[\tilde{\mathbf{I}}\left(\tilde{\boldsymbol{\theta}}\right)\right]_{\Re\Re}\right. \\ \left. +\left[\tilde{\mathbf{I}}\left(\tilde{\boldsymbol{\theta}}\right)\right]_{\Im\Im}\right)+\frac{j}{4}\left(\left[\tilde{\mathbf{I}}\left(\tilde{\boldsymbol{\theta}}\right)\right]_{\Re\Im}-\left[\tilde{\mathbf{I}}\left(\tilde{\boldsymbol{\theta}}\right)\right]_{\Im\Re}\right).
\label{eq:cr_real2complex}
\end{multline}

The trace of $\mathbf{I}\left(\mathbf{h}\right)^{-1}$ will be used in our simulations as the variance of the estimation error.

\subsection{Algorithms for Channel Estimation and Data Equalization}\label{sec:algorithms}
In this section, we will introduce some algorithms that can be used to solve the models (\ref{eq:channel_estimation_model}),  (\ref{eq:data_equalization_model}), (\ref{eq:channel_estimation_model_sc}) and (\ref{eq:data_equalization_model_sc}). Again, we express the model generically as:
\begin{equation}
\mathbf{y}=\mathcal{Q}\left(\mathbf{A}\mathbf{h}+\mathbf{w}\right),
\label{eq:model_column}
\end{equation}
where we denote the dimensions of $\mathbf{A}$ as $K\times P$.

There are two broad classes of algorithms. The joint estimation algorithms take the whole model into account. They provide near optimal results, but are iterative and need to operate with large matrices. The subcarrier-wise estimation algorithms linearize the problem. This leads to a loss in performance, but enables independent processing of each subcarrier, drastically reducing complexity in the OFDM case.
\subsection{Expectation Maximization (EM) Algorithm}
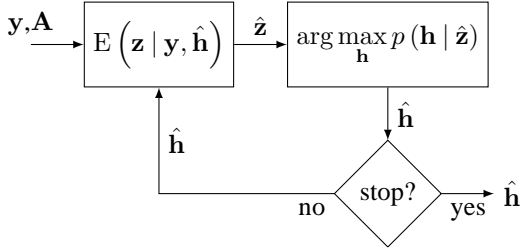
\begin{figure}[!t]\centering
\begin{tikzpicture}[
	node distance=2em,
	block/.style={
		rectangle,
		minimum width=5em,
		minimum height=3.21em,
		draw
	},
	diamond/.style={
		dspdiamond,
		minimum width=4em,
		minimum height=4em,
		draw
	}
]
\node (y) {$\mathbf{y}$,$\mathbf{A}$};
\node (E) [block, right=of y.south] {$\mathrm{E}\left(\mathbf{z}\given{\mathbf{y},\hat{\mathbf{h}}}\right)$};
\node (M) [block, right=of E] {$\arg\max\limits_{\mathbf{h}} p\left(\mathbf{h}\given{\hat{\mathbf{z}}}\right)$};
\node (stop) [diamond, below=of M] {stop?};
\node at (stop.east) [anchor=north west] {yes};
\node at (stop.west) [anchor=north east] {no};

\draw [-latex] (y.south) -- (E);
\draw [-latex] (E) -- node[above] {$\hat{\mathbf{z}}$} (M);
\draw [-latex] (M) -- node[right] {$\hat{\mathbf{h}}$} (stop);
\draw [-latex] (stop) -| node[right, near end] {$\hat{\mathbf{h}}$} (E);
\draw [-latex] (stop.east) -- ++ (\nodedistance, 0em) node [right] {$\hat{\mathbf{h}}$};

\end{tikzpicture}
\caption{Expectation Maximization-MMSE (EM-MMSE) algorithm}
\label{fig:em}
\end{figure}

The Expectation Maximization (EM) approach \cite{dempster_EM} is iterative and alternately applies two steps at each iteration $i$:

\subsubsection{Expectation step:} obtain the expected value of the unquantized observations $\mathbf{z}=\mathbf{A}\mathbf{h}+\mathbf{w}$, given the quantized output $\mathbf{y}$ and the current estimate $\hat{\mathbf{h}}^{(i-1)}$:
\begin{equation}
\hat{\mathbf{z}}^{(i)}=\mathrm{E}\left(\mathbf{z}\given{\mathbf{y}, \hat{\mathbf{h}}^{(i-1)}}\right).
\end{equation}
A closed form expression for this expectation is derived in Appendix~\ref{sec:em}. The result is:
\begin{equation}
\hat{\mathbf{z}}^{(i)}=\mathbf{A}\hat{\mathbf{h}}^{(i-1)}+\hat{\mathbf{w}},
\label{eq:em_e}
\end{equation}
where the components of $\hat{\mathbf{w}}$ are given by:
\begin{equation}
\hat{w}_k=\frac{\sigma_k}{\sqrt{2}}\left(\frac{\Real{y_k}\phi\left(\tilde{\eta}_k\right)}{\mathrm{\Phi}\left(\tilde{\eta}_k\right)}+j\frac{\Imag{y_k}\phi\left(\overline{\eta}_k\right)}{\mathrm{\Phi}\left(\overline{\eta}_k\right)}\right),
\label{eq:E_z_yh_1b}
\end{equation}
where
\begin{equation}
\tilde{\eta}_k=\frac{\Real{y_k}\Real{\sum\limits_{p=1}^{P}a_{kp}h_p}}{\sigma_k/\sqrt{2}};\ \overline{\eta}_k=\frac{\Imag{y_k}\Imag{\sum\limits_{p=1}^{P}a_{kp}h_p}}{\sigma_k/\sqrt{2}}.
\label{eq:em_eta_mu}
\end{equation}

\subsubsection{Maximization Step}
In the maximization step, the obtained $\hat{\mathbf{z}}^{(i)}\triangleq\mathrm{E}\left\{\mathbf{z}\given{\mathbf{y},\hat{\mathbf{h}}}^{(i-1)}\right\}$ (\ref{eq:E_z_yh_1b}) is used as observation vector in an unquantized problem:
\begin{equation}
\hat{\mathbf{z}}^{(i)}=\mathbf{A}\mathbf{h}+\mathbf{w}.
\end{equation}
In~\cite{mo_mimo_1b_chest}, a maximum likelihood (ML) estimator is used, but we propose an MMSE estimator as an alternative:
\begin{equation}
\hat{\mathbf{h}}^{(i)}=\left(\mathbf{A}^H\mathbf{R}_{\mathbf{w}\mathbf{w}}^{-1}\mathbf{A}+\mathbf{R}_{\mathbf{h}\mathbf{h}}^{-1}\right)^{-1}\mathbf{A}^H\mathbf{R}_{\mathbf{w}\mathbf{w}}^{-1}\hat{\mathbf{z}}^{(i)};
\label{eq:em_mmse_h}
\end{equation}
This solution (EM-MMSE) gives better performance, as it takes into account prior information. The program flow of the full EM-MMSE method is graphically depicted in Fig.~\ref{fig:em}, and an implementation in pseudo-code is given in Algorithm~\ref{algo:EM-ML}. The initialization of $\hat{\mathbf{h}}^{(0)}$ in the first step is the Least Squares solution that ignores quantization, which provides an acceptable starting point for the optimization.

\begin{algorithm}
\caption{Expectation Maximization (EM)}
\label{algo:EM-ML}
\begin{algorithmic}
\REQUIRE $\mathbf{A}$, $\mathbf{y}$
\STATE \textbf{Initialize:} $\hat{\mathbf{h}}^{(0)}=\left(\mathbf{A}^H\mathbf{A}\right)^{-1}\mathbf{A}^H\mathbf{y}$, $i=1$
\WHILE{$i\le i_{\max}$ and  $\left\|\hat{\mathbf{h}}^{(i)}-\hat{\mathbf{h}}^{(i-1)}\right\|_F^2\ge\epsilon\left\|\mathbf{h}^{(i)}\right\|_F^2$}
\STATE $\hat{\mathbf{z}}=\mathrm{E}\left(\mathbf{z}\given{\mathbf{y},\hat{\mathbf{h}}^{(i-1)}}\right)$ from (\ref{eq:em_e})
\STATE $\hat{\mathbf{h}}^{(i)}$ from (\ref{eq:em_mmse_h})
\STATE $i=i+1$
\ENDWHILE
\ENSURE $\hat{\mathbf{h}}^{(i)}$
\end{algorithmic}
\end{algorithm}

\subsection{Generalized Approximate Message Passing (GAMP)}
\begin{figure}[htbp]\centering

\begin{tikzpicture}[
	node distance=2.5em,
	block/.style={
		minimum width=3.5em,
		minimum height=2.47em,
		rectangle,
		draw
	}
]
\node (px_q) [block] {$p_{\mathbf{x}}$};
\node (A) [block, right=of px_q] {$\mathbf{A}$};
\node (py_z) [block, right=of A] {$p_{\mathbf{y}\given{\mathbf{z}}}$};
\node (y) [coordinate, right=of py_z] {};

\path[arrows={-latex}]
	(px_q) edge node[above, fill=black!30!white] {$\mathbf{x}$} (A)
	(A) edge node[above, fill=black!30!white] {$\mathbf{z}$} (py_z)
	(py_z) edge (y);
	
\node at (y) [anchor=south west] {$\mathbf{y}$};
\end{tikzpicture}

\caption{Problem formulation of GAMP: the unknown signals to estimate are shaded in gray}
\label{fig:gamp}
\end{figure}
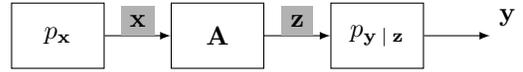

The Generalized Approximate Message Passing method, developed in~\cite{rangan_GAMP}, can also be applied to the quantized estimation problem. This is a very general method for estimation in coupled nonlinear channels with the structure depicted in Figure~\ref{fig:gamp}. An input signal $\mathbf{x}\inC{P}$ with known prior $p_{\mathbf{x}}\left(\mathbf{x}\right)$ goes through a linear transformation $\mathbf{z}=\mathbf{A}\mathbf{x}\inC{K}$, and then through the output channel $p_{\mathbf{y}\given{\mathbf{z}}}$, giving the observed output $\mathbf{y}$. GAMP estimates the input signal $\mathbf{x}$ and the intermediate signal $\mathbf{z}$ from the knowledge of $\mathbf{A}\inC{K\times P}$, $\mathbf{y}$, $p_{\mathbf{x}}$, and $p_{\mathbf{y}\given{\mathbf{z}}}$ by using a loopy belief propagation approach. The details of the algorithm are explained in~\cite{rangan_GAMP},
 and a listing is provided in Algorithm~\ref{algo:GAMP} (where $\odot$ denotes elementwise product).
\begin{algorithm}[htbp]
\caption{Generalized Approximate Message Passing (GAMP)}
\label{algo:GAMP}
\begin{algorithmic}
\REQUIRE $\mathbf{A}$, $\mathbf{y}$, $p_{\mathbf{x}}$, and $p_{\mathbf{y}\given{\mathbf{z}}}$
\STATE \textbf{Compute:} $\mathbf{A}_2=\left|\mathbf{A}\right|^2$ (elementwise)
\STATE \textbf{Initialize:} $i=0$, $\mathbf{s}=\mathbf{0}_{P\times 1}$, $\mathbf{x}$, $\boldsymbol{\tau}^x$
\WHILE{$i<i_{\max}$ \textbf{and} $\left\|\mathbf{x}-\mathbf{x}(i-1)\right\|_2^2\ge\epsilon\left\|\mathbf{x}\right\|_2^2$}
\STATE $\mathbf{x}_{\mathrm{old}}=\mathbf{x}$
\STATE \textbf{Output linear step:}
\STATE $\boldsymbol{\tau}^p=\mathbf{A}_2\boldsymbol{\tau}^x$
\STATE $\mathbf{p}=\mathbf{A}\mathbf{x}-\frac{1}{2}\boldsymbol{\tau}^p\odot\mathbf{s}$
\STATE $\mathbf{z}=\mathbf{A}\mathbf{x}$
\STATE \textbf{Output nonlinear step:}
\FOR {$\ell=1:K$}
\STATE $s_\ell=g_{out}\left(i, p_\ell,y_\ell, \tau_\ell^p \right)$
\STATE $\tau_\ell^s=-\frac{\partial}{\partial p}g_{out}\left(i, p_\ell,y_\ell, \tau_\ell^p \right)$
\ENDFOR
\STATE \textbf{Input linear step:}
\STATE $\boldsymbol{\tau}^r=4/\left(\mathbf{A}_2^H\boldsymbol{\tau}^s\right)$ (elementwise inverse)
\STATE $\mathbf{r}=\mathbf{x}+\frac{1}{2}\boldsymbol{\tau}^r\odot \left(\mathbf{A}^T\mathbf{s}\right)$
\STATE \textbf{Input nonlinear step:}
\FOR{$\ell=1:P$}
\STATE $x_\ell=g_{in}\left(i, r_\ell, \tau_\ell^r \right)$
\STATE $\tau_\ell^x=\tau^r\frac{\partial}{\partial r}g_{in}\left(i, r_\ell \tau_\ell^r \right)$
\ENDFOR
\STATE \textbf{Increment loop index:} $i=i+1$
\ENDWHILE
\ENSURE $\mathbf{x}$, $\mathbf{z}$ 
\end{algorithmic}
\end{algorithm}

The scalar estimation functions $g_{in}$ and $g_{out}$ for our quantized estimation problems, as well as their derivatives, are given in the following (we provide a detailed derivation of these results in Appendix~\ref{app:gamp_steps}):
\begin{itemize}
\item Input nonlinear step for Gaussian input (estimation):
\begin{equation}
g_{in}\left(i, r, \tau^r\right)=\frac{\sigma_x^2}{\sigma_x^2+\tau^r}r,
\label{eq:gamp_g_in}
\end{equation}
\begin{equation}
\tau^r\frac{\partial}{\partial r}g_{in}\left(i, r, \tau^r\right)=\frac{\sigma_x^2\tau^r}{\sigma_x^2+\tau^r},
\label{eq:gamp_dg_in}
\end{equation}
\item Input nonlinear step for discrete input (equalization):
\begin{equation}
g_{in}\left(i, r, \tau^r\right)=\frac{\sum\limits_{a=1}^{A} P_a\overline{x}_a e^{-\frac{\left|r-\overline{x}_a\right|^2}{\tau^r}}}{\sum\limits_{a=1}^{A}P_a e^{-\frac{\left|r-\overline{x}_a\right|^2}{\tau^r}}},
\label{eq:gamp_g_in_c}
\end{equation}
\begin{multline}
\tau^r\frac{\partial}{\partial r}g_{in}\left(i, r, \tau^r\right) \\ =\frac{\sum\limits_{a=1}^{A} P_a\left|\overline{x}_a\right|^2 e^{-\frac{\left|r-\overline{x}_a\right|^2}{\tau^r}}}{\sum\limits_{a=1}^{A}P_a e^{-\frac{\left|r-\overline{x}_a\right|^2}{\tau^r}}}-\left|g_{in}\left(i, r, \tau^r\right)\right|^2,
\label{eq:gamp_dg_in_c}
\end{multline}
where $\overline{x}_a, a\inset{1}{A}$ are the constellation symbols, and $P_a$ are their corresponding probabilities.
\item Output nonlinear step for 1-bit quantization: for this step, the functions are applied separately to the real and imaginary parts:
\begin{equation}
g_{out,\Re}\left(i, p, y, \tau^p\right)=\frac{y\sqrt{2}}{\sqrt{\sigma_w^2+\tau^p}}\frac{\phi\left(\eta\right)}{\mathrm{\Phi}\left(\eta\right)}
\label{eq:gamp_g_out}
\end{equation}
\begin{equation}
-\frac{\partial}{\partial p}g_{out,\Re}\left(i, p, y, \tau^p\right)=\frac{2}{\sigma_w^2+\tau^p}\left(\eta\frac{\phi\left(\eta\right)}{\mathrm{\Phi}\left(\eta\right)}+\frac{\phi\left(\eta\right)^2}{\mathrm{\Phi}\left(\eta\right)^2}\right),
\label{eq:gamp_dg_out}
\end{equation}
where
\begin{equation}
\eta=\frac{yp\sqrt{2}}{\sqrt{\sigma_w^2+\tau^p}}.
\end{equation}
The complex-valued output nonlinear step is given by:
\begin{multline}
g_{out}\left(i, p, y, \tau^p\right) \\ =g_{out,\Re}\left(i, p_{\Re}, y_{\Re}, \tau^p\right)+jg_{out,\Re}\left(i, p_{\Im}, y_{\Im}, \tau^p\right),
\end{multline}
\begin{multline}
-\frac{\partial}{\partial p}g_{out,\Re}\left(i, p, y, \tau^p\right) \\ =-\frac{\partial}{\partial p}g_{out,\Re}\left(i, p_{\Re}, y_{\Re}, \tau^p\right)-\frac{\partial}{\partial p}g_{out,\Re}\left(i, p_{\Im}, y_{\Im}, \tau^p\right),
\end{multline}
with $p_\Re\triangleq\Real{p}$, $p_\Im\triangleq{\Imag{p}}$.
\end{itemize}

\subsection{Subcarrier-Wise Estimation with Bussgang Theorem}
Both EM and GAMP have very high complexity, and are not practical for Massive MIMO scenarios, or for high numbers of subcarriers. In this section, a linear estimator based on the Bussgang theorem is proposed. This theorem~\cite{bussgang_theorem} states that a nonlinear distortion of a Gaussian signal can be expressed as a linear transformation plus uncorrelated noise.

Let us consider the real-valued version of the problem, $\tilde{\mathbf{y}}=\mathcal{Q}\left(\tilde{\mathbf{A}}\tilde{\mathbf{h}}+\tilde{\mathbf{w}}\right)$, as given by (\ref{eq:make_real}). By the Bussgang theorem, (\ref{eq:cr_real_valued}) can be modeled as:
\begin{equation}
\tilde{\mathbf{y}}=\tilde{\mathbf{K}}\tilde{\mathbf{z}}+\tilde{\mathbf{e}},
\label{eq:bussgang_model}
\end{equation}
and $\tilde{\mathbf{K}}\inR{2K\times 2K}$ can be chosen such that:
\begin{equation}
\mathrm{E}\left\{\tilde{\mathbf{z}}\tilde{\mathbf{e}}^H\right\}=\mathbf{0}.
\end{equation}
Using $\tilde{\mathbf{e}}=\tilde{\mathbf{y}}-\tilde{\mathbf{K}}\tilde{\mathbf{z}}$, it is easy to derive $\tilde{\mathbf{K}}$ and the covariance matrix of the quantization noise, $\mathbf{R}_{\tilde{\mathbf{e}}\tilde{\mathbf{e}}}$:
\begin{equation}
\tilde{\mathbf{K}}=\mathbf{R}_{\tilde{\mathbf{y}}\tilde{\mathbf{z}}}\mathbf{R}_{\tilde{\mathbf{z}}\tilde{\mathbf{z}}}^{-1};
\label{eq:bussgang_K}
\end{equation}
\begin{equation}
\mathbf{R}_{\tilde{\mathbf{e}}\tilde{\mathbf{e}}}=\mathbf{R}_{\tilde{\mathbf{y}}\tilde{\mathbf{y}}}-\mathbf{R}_{\tilde{\mathbf{y}}\tilde{\mathbf{z}}}\mathbf{R}_{\tilde{\mathbf{z}}\tilde{\mathbf{z}}}^{-1}\mathbf{R}_{\tilde{\mathbf{z}}\tilde{\mathbf{y}}}.
\label{eq:bussgang_Ree}
\end{equation}
For our problem, we have:
\begin{equation}
\mathbf{R}_{\tilde{\mathbf{z}}\tilde{\mathbf{z}}}=\tilde{\mathbf{A}}\mathbf{R}_{\tilde{\mathbf{h}}\tilde{\mathbf{h}}}\tilde{\mathbf{A}}^H+\mathbf{R}_{\tilde{\mathbf{w}}\tilde{\mathbf{w}}}.
\end{equation}
The Bussgang gain and noise covariance are easily adapted from the results in~\cite{wendler_bussgang}:
\begin{equation}
\tilde{\mathbf{K}}=\sqrt{\frac{2}{\pi}}\mathrm{diag}\left\{\mathbf{R}_{\tilde{\mathbf{z}}\tilde{\mathbf{z}}}\right\}^{-1/2};
\end{equation}
\begin{equation}
\mathbf{R}_{\tilde{\mathbf{e}}\tilde{\mathbf{e}}}=\mathbf{R}_{\tilde{\mathbf{y}}\tilde{\mathbf{y}}}-\frac{2}{\pi}\mathrm{diag}\left\{\mathbf{R}_{\tilde{\mathbf{z}}\tilde{\mathbf{z}}}\right\}^{-1/2}\mathbf{R}_{\tilde{\mathbf{z}}\tilde{\mathbf{z}}}\mathrm{diag}\left\{\mathbf{R}_{\tilde{\mathbf{z}}\tilde{\mathbf{z}}}\right\}^{-1/2},
\end{equation}
where
\begin{equation}
\mathbf{R}_{\tilde{\mathbf{y}}\tilde{\mathbf{y}}}=\frac{2}{\pi}\arcsin\left(\mathrm{diag}\left\{\mathbf{R}_{\tilde{\mathbf{z}}\tilde{\mathbf{z}}}\right\}^{-1/2}\mathbf{R}_{\tilde{\mathbf{z}}\tilde{\mathbf{z}}}\mathrm{diag}\left\{\mathbf{R}_{\tilde{\mathbf{z}}\tilde{\mathbf{z}}}\right\}^{-1/2}\right).
\label{eq:bussgang_Ryy}
\end{equation}

Using (\ref{eq:bussgang_model}), we can now model our quantized system (\ref{eq:model_column}) as an unquantized one:
\begin{equation}
\tilde{\mathbf{y}}=\tilde{\mathbf{B}}\tilde{\mathbf{h}}+\tilde{\boldsymbol{\eta}},
\label{eq:bussgang_end}
\end{equation}
where $\tilde{\mathbf{B}}=\tilde{\mathbf{K}}\tilde{\mathbf{A}}$, and $\mathbf{R}_{\tilde{\boldsymbol{\eta}}\tilde{\boldsymbol{\eta}}}=\tilde{\mathbf{K}}\mathbf{R}_{\tilde{\mathbf{w}}\tilde{\mathbf{w}}}\tilde{\mathbf{K}}^H+\mathbf{R}_{\tilde{\mathbf{e}}\tilde{\mathbf{e}}}$. Note that the quantization noise $\tilde{\mathbf{e}}$ is \textit{not} Gaussian, and therefore this approach is suboptimal. If $\mathbf{R}_{\tilde{\mathbf{z}}\tilde{\mathbf{z}}}$ is assumed to be diagonal (which holds if $\mathbf{R}_{\tilde{\mathbf{h}}\tilde{\mathbf{h}}}$ is diagonal, the pilots are orthogonal and the number of transmit antennas is large), the problem can be decoupled and $\mathbf{R}_{\tilde{\boldsymbol{\eta}}\tilde{\boldsymbol{\eta}}}$ reduces to:
\begin{equation}
\mathbf{R}_{\tilde{\boldsymbol{\eta}}\tilde{\boldsymbol{\eta}}}=\frac{2}{\pi}\mathrm{diag}\left\{\mathbf{R}_{\tilde{\mathbf{z}}\tilde{\mathbf{z}}}\right\}^{-1}\mathrm{diag}\left\{\mathbf{R}_{\tilde{\mathbf{w}}\tilde{\mathbf{w}}}\right\}+2\left(1-\frac{2}{\pi}\right)\mathbf{I}_M.
\end{equation}

If, additionally, $\mathbf{R}_{\tilde{\mathbf{z}}\tilde{\mathbf{z}}}$ and $\mathbf{R}_{\tilde{\mathbf{w}}\tilde{\mathbf{w}}}$ are scaled identities (which, if diagonality is already assumed, only requires that the noise and pilots do not change their variance over time), then the problem simplifies even further. In this case, the Bussgang decomposition reduces to a scalar factor and i.i.d. noise, and from (\ref{eq:model}), we have:
\begin{equation}
\mathbf{Y}_{n_r}=\mathcal{Q}\left(\sum_{n_t=1}^{N_t}\mathbf{F}^H\mathbf{\Lambda}_{n_rn_t}\mathbf{X}_{n_t}+\mathbf{W}_{n_r}\right),
\end{equation}
\begin{equation}
\mathbf{Y}_{n_r}=k\sum_{n_t=1}^{N_t}\mathbf{F}^H\mathbf{\Lambda}_{n_rn_t}\mathbf{X}_{n_t}+\tilde{\mathbf{W}}_{n_r},
\end{equation}
with $k=\frac{1}{\sigma_z}\sqrt{\frac{2}{\pi}}$, and $\sigma_{\tilde{w}}^2=\frac{2}{\pi}\frac{\sigma_w^2}{\sigma_z^2}+\left(1-\frac{2}{\pi}\right)$. This allows to use standard OFDM techniques: apply an FFT to $\mathbf{Y}$, and then estimate $\mathbf{H}$ subcarrier-wise:
\begin{equation}
\bm{Y}_{n_r}=\mathbf{F}\mathbf{Y}_{n_r}=k\sum_{n_t=1}^{N_t}\mathrm{diag}\left\{\mathcal{H}_{n_rn_t\cdot}\right\}\mathbf{X}_{n_t}+\btm{W}_{n_r},
\end{equation}
\begin{equation}
\bm{Y}_{ n}=k\mathbf{H}_{ n}\mathbf{X}_{ n}+\btm{W}_{ n},
\end{equation}
where $\bm{Y}_{ n}\inC{N_r\times T}$ , $\mathbf{H}_{ n}\inC{N_r\times N_t}$, $\mathbf{X}_{ n}\inC{N_t\times T}$, and $\btm{W}_{ n}\inC{N_r\times T}$ are respectively the frequency-domain observations, channel, pilots and noise at subcarrier $n$. Then, the frequency-domain channel estimation at each subcarrier can be done, for example, using a linear MMSE algorithm:
\begin{equation}
\bhm{H}_{ n}=\frac{1}{k}\bm{Y}_{ n}\mathbf{X}_{ n}^H
\left(\mathbf{X}_{ n}\mathbf{X}_{ n}^H+\frac{\sigma_{\tilde{w}}^2}{\sigma_h^2}\mathbf{I}_{N_t}\right)^{-1}.
\label{eq:bussgang_iid_mmse}
\end{equation}

\subsection{Computational Complexity}\label{sec:complexity}
In this section, we compare the computational complexity (number of complex multiplications) of the presented algorithms. We define $K$ and $P$ such that $\mathbf{A}\inC{K\times P}$ in (\ref{eq:model_column}), and $R=N_r$ for estimation and $R=M$ for equalization.

\subsubsection{Computational Complexity of EM}
\begin{itemize}
\item \textbf{Expectation step:} $\mathrm{E}\left\{\mathbf{z}\given{\mathbf{y},\hat{\mathbf{h}}^{(i)}}\right\}$. This amounts to $R$ computations of (\ref{eq:em_e}), each one having a complexity dominated by the product $\mathbf{A}\hat{\mathbf{h}}$, yielding $\mathcal{O}\left(KPR\right)$.
\item \textbf{Maximization step:} this multiplies $\mathbf{B}=\left(\mathbf{A}^H\mathbf{R}_{\mathbf{w}\mathbf{w}}\mathbf{A}+\mathbf{R}_{\mathbf{h}\mathbf{h}}\right)^{-1}\mathbf{A}\mathbf{R}_{\mathbf{w}\mathbf{w}}\inC{P\times K}$ by the expectation $\hat{\mathbf{z}}\inC{K\times 1}$.  Note that $\mathbf{B}$ only needs to be calculated once in each channel coherence period. The maximization step amounts to a matrix-vector multiplication $\mathbf{B}\mathbf{y}$, which is done $R$ times. The complexity of this step is: $\mathcal{O}\left(KPR\right)$.
\end{itemize}
These two steps are done for $I$ iterations, until the algorithm converges. This results in an overall complexity of EM of:
\begin{equation} T_{\mathrm{EM}}=\mathcal{O}\left(2IKPR\right).
\end{equation}

\subsubsection{Computational Complexity of GAMP}
The most computationally expensive step of each iteration of GAMP are two matrix-vector multiplications involving $\mathbf{A}$ and another two involving $\mathbf{A}_2$. All of them have complexity $KP$. Taking into account $R$ runs with $I$ iterations:
\begin{equation}
T_{\mathrm{GAMP-MMSE}}=\mathcal{O}\left(4IKPR\right).
\end{equation}

\subsubsection{Computational Complexity of the Bussgang estimator}
In the single-carrier case, the linear estimator needs to compute an MMSE solution with the whole matrix $\tilde{\mathbf{K}}\tilde{\mathbf{A}}\inR{2K\times 2P}$ and observation $\tilde{\mathbf{Y}}\inR{2K\times 2P}$. Again, note that the computation of the matrices $\tilde{\mathbf{K}}\tilde{\mathbf{A}}$ and $\mathbf{R}_{\tilde{\boldsymbol{\eta}}\tilde{\boldsymbol{\eta}}}$ only needs to be performed once every channel realization, and the same applies to the MMSE multiplier matrix $\tilde{\mathbf{G}}=\left(\tilde{\mathbf{A}}^H\tilde{\mathbf{K}}^H\mathbf{R}_{\tilde{\boldsymbol{\eta}}\tilde{\boldsymbol{\eta}}}^{-1}\tilde{\mathbf{K}}\tilde{\mathbf{A}}+\mathbf{R}_{\tilde{\mathbf{h}}\tilde{\mathbf{h}}}\right)^{-1}\tilde{\mathbf{A}}^H\tilde{\mathbf{K}}^H$ The complexity of the Bussgang estimator then reduces to a real-valued matrix-vector multiplication of $\tilde{\mathbf{G}}\inR{2P\times 2K}$ with $\tilde{\mathbf{y}}$, which is done $R$ times. An additional factor of $1/4$ is applied to the complexity because we are measuring it in terms of complex-valued multiplications:

\begin{equation}
T_{\mathrm{Buss.-SC}}=\mathcal{O}\left(KPR\right).
\end{equation}

In an OFDM system, the Bussgang estimator allows subcarrier-wise equalization, which amounts to $N$ MMSE calculations of (\ref{eq:bussgang_iid_mmse}), where again the matrix inversion only needs to be performed once per channel realization.

All the complexity results for channel estimation and equalization are summarized in Table~\ref{tab:complexity}, where MC stands for multi-carrier (OFDM).

\begin{table}[htbp]\centering
\caption{Computational complexity of the presented algorithms}\label{tab:complexity}
\begin{tabular}{| c | c | c |}
\hline
\textbf{Algorithm} & \textbf{Estimation} & \textbf{Equalization}
\\
\hline
EM & $2IN_rN_tNLT$ & $2IN_rN_tN^2M$ \\ GAMP & $4IN_rN_tNLT$ & $4IN_rN_tN^2M$ \\ Buss. (SC) & $N_rN_tNLT$ & $N_rN_tN^2M$ \\ Buss. (MC) & $N_rN_tNT$ & $N_rN_tNM$ \\
\hline
\end{tabular}
\end{table}

\section{Simulation Results}\label{sec:results}
This section presents simulation results that validate the channel estimation and data equalization models proposed in this paper, and compare the presented algorithms. 

All experiments simulate a system with block size of $N=32$ and QPSK modulation, unless otherwise stated. The noise is AWGN with variance $\sigma_w^2=1$. A punctured convolutional code (CC) of rate $3/4$ is used in all systems for the transmission of the data bits. The channel length is $L=4$, and its taps are i.i.d. Gaussian in all experiments except for the last. The signal to noise ratio (SNR) is defined as $P_t/N_t$, where $P_t$ is the transmitted power. The Normalized Mean Square Error (NMSE) of the channel estimate is defined as:
\begin{equation}
\mathrm{NMSE}_{\mathbf{H}}=\frac{1}{N_rN_tL\sigma_h^2}\sum_{n_r=1}^{N_r}\sum_{n_t=1}^{N_t}\left\|\hat{\mathbf{h}}_{n_rn_t}-\mathbf{h}_{n_rn_t}\right\|_2^2.
\end{equation}
The coherence time of the channels is set to $M=64$ symbols, and the results are averaged over 4096 channel realizations, which corresponds to $2.5\cdot{10}^7$ data bits.

\subsection{Comparison of Algorithms for Channel Estimation}
\begin{figure}[!t]\centering

%
%
\definecolor{color_em}{rgb}{0.00000,0.50000,0.00000}%
\definecolor{color_gamp}{rgb}{1.00000,0.00000,0.00000}%
\definecolor{color_bussgang}{rgb}{0.00000,0.00000,1.00000}%
\definecolor{color_ignoring}{rgb}{1.00000,0.500000,0.00000}%
\begin{tikzpicture}

\begin{axis}[%
width=0.95092\figurewidth,
height=\figureheight,
at={(0\figurewidth,0\figureheight)},
scale only axis,
every outer x axis line/.append style={black},
every x tick label/.append style={font=\color{black}},
xmin=-9,
xmax=3,
xtick={-9,-7,-5,-3,-1,1,3},
xlabel={SNR ($\mathrm{dB}$)},
every outer y axis line/.append style={black},
every y tick label/.append style={font=\color{black}},
ymode=log,
ymin=8e-3,
ymax=0.2,
yminorticks=true,
xmajorgrids,
ymajorgrids,
yminorgrids,
ylabel={$\mathrm{NMSE}_{\mathbf{H}}$},
axis x line*=bottom,
axis y line*=left,
legend columns=3,
legend style={at={(0.03,0.03)},anchor=south west,legend cell align=left,align=left,draw=black,row sep=0em}
]
\addplot [color=color_em,very thick,mark=square,mark options={solid}]
  table[row sep=crcr]{%
-9	0.127668515553741\\
-7	0.100527268928749\\
-5	0.0816495759634947\\
-3	0.0687466498238487\\
-1	0.0598925890234923\\
1	0.0543192959811993\\
3	0.0516923317724684\\
};
\addlegendentry{EM-MC};

\addplot [color=color_em, very thick, dashed,mark=square,mark options={solid}]
  table[row sep=crcr]{%
-9	0.114548634284765\\
-7	0.0910898868492493\\
-5	0.0753759393655001\\
-3	0.0658307792864963\\
-1	0.060262950563091\\
1	0.0578467017517079\\
3	0.0592905625200221\\
};
\addlegendentry{EM-SC};

\addplot [color=color_gamp,very thick,mark=triangle,mark options={solid,rotate=180}]
  table[row sep=crcr]{%
-9	0.121126041654497\\
-7	0.0955712729833078\\
-5	0.0776094722781235\\
-3	0.0652194197036269\\
-1	0.0566115638742533\\
1	0.0508895717395899\\
3	0.0475030156702889\\
};
\addlegendentry{GAMP-MC};

\addplot [color=color_gamp, very thick, dashed,mark=triangle,mark options={solid,rotate=180}]
  table[row sep=crcr]{%
-9	0.109145700888901\\
-7	0.0869558721417493\\
-5	0.0720113617058735\\
-3	0.0629106767568883\\
-1	0.0575960523988182\\
1	0.0552658007855458\\
3	0.0557750285246024\\
};
\addlegendentry{GAMP-SC};

\addplot [color=color_bussgang,very thick,mark=o,mark options={solid}]
  table[row sep=crcr]{%
-9	0.130883920174371\\
-7	0.109439870302038\\
-5	0.0959444812881122\\
-3	0.0877277364155753\\
-1	0.0828184184292095\\
1	0.0799228871977239\\
3	0.0783041981570244\\
};
\addlegendentry{Buss.-MC};

\addplot [color=color_bussgang, very thick, dashed,mark=o,mark options={solid}]
  table[row sep=crcr]{%
-9	0.123218424847908\\
-7	0.107637250300615\\
-5	0.099820044791266\\
-3	0.0972574202479388\\
-1	0.0975118943188237\\
1	0.0993653548546519\\
3	0.101930852550682\\
};
\addlegendentry{Buss.-SC};

\addplot [color=color_ignoring, very thick,mark=triangle,mark options={solid}]
  table[row sep=crcr]{%
-9	0.155625964146919\\
-7	0.137367745870853\\
-5	0.126866091990706\\
-3	0.121046423130573\\
-1	0.117982451262729\\
1	0.116383115785886\\
3	0.115575656840789\\
};
\addlegendentry{Ign.-MC};

\addplot [color=color_ignoring, very thick, dashed,mark=triangle,mark options={solid}]
  table[row sep=crcr]{%
-9	0.129565590258492\\
-7	0.113231452724633\\
-5	0.105328739397601\\
-3	0.103048901798082\\
-1	0.103648475588818\\
1	0.105761448121341\\
3	0.108394433333413\\
};
\addlegendentry{Ign.-SC};

\addplot [color=black, very thick]
  table[row sep=crcr]{%
-9	0.134083433071146\\
-7	0.100360689897333\\
-5	0.0770057445003019\\
-3	0.0603786498832556\\
-1	0.0482106440276266\\
1	0.0390962224400923\\
3	0.0321591695830434\\
};
\addlegendentry{CRLB-MC};

\addplot [color=black,very thick,dashed]
  table[row sep=crcr]{%
-9	0.116725542764022\\
-7	0.0869840608422458\\
-5	0.0667581733730772\\
-3	0.0526025966180389\\
-1	0.0424363072747306\\
1	0.0350066203768541\\
3	0.0295627514413588\\
};
\addlegendentry{CRLB-SC};

\end{axis}
\end{tikzpicture}%
\caption{NMSE comparison of channel estimation techniques with EM equalization ($N_r\times N_t=10\times 2$, $N=32$ subcarriers, $T=4$ pilot blocks)}
\label{fig:paper_methods_chest_nmse}
\end{figure}
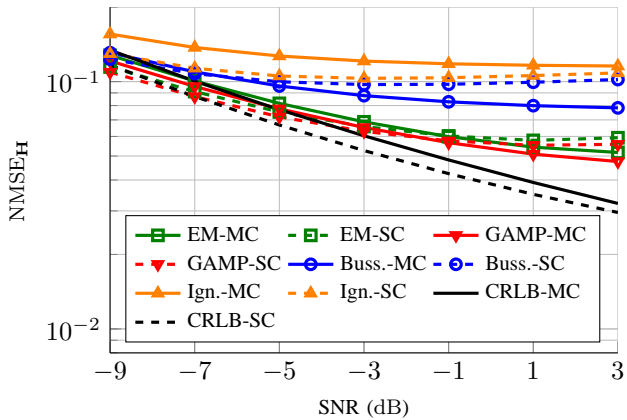

In the first experiment, systems with $N_r=10$ receive and $N_t=2$ transmit antennas were simulated. Using $T=4$ orthogonal pilot blocks, different methods for channel estimation were compared, both in the OFDM multi-carrier (MC) case and in the single carrier (SC) one. The compared methods are EM, GAMP, Bussgang and Ignoring (which performs linear estimation as if there were no quantizer). All the systems use Expectation Maximization (EM) for equalization. Fig.~\ref{fig:paper_methods_chest_nmse} shows the estimation error. The Cram\'er-Rao bound derived in (\ref{eq:cr_bound}) is also given. The bound is tight at low SNR, but there might still be room for improvement in the high SNR region. 

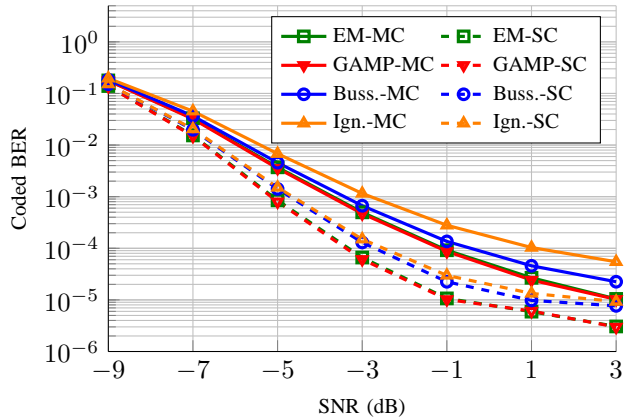
\begin{figure}[!t]\centering
%
%
\definecolor{color_em}{rgb}{0.00000,0.50000,0.00000}%
\definecolor{color_gamp}{rgb}{1.00000,0.00000,0.00000}%
\definecolor{color_bussgang}{rgb}{0.00000,0.00000,1.00000}%
\definecolor{color_ignoring}{rgb}{1.00000,0.500000,0.00000}%
\begin{tikzpicture}

\begin{axis}[%
width=0.95092\figurewidth,
height=\figureheight,
at={(0\figurewidth,0\figureheight)},
scale only axis,
every outer x axis line/.append style={black},
every x tick label/.append style={font=\color{black}},
xmin=-9,
xmax=3,
xtick={-9,-7,-5,-3,-1,1,3},
xlabel={SNR (dB)},
every outer y axis line/.append style={black},
every y tick label/.append style={font=\color{black}},
ymode=log,
ymin=1e-6,
ymax=5,
ytick={1e-6,1e-5,1e-4,1e-3,1e-2,1e-1,1},
yminorticks=true,
xmajorgrids,
ymajorgrids,
yminorgrids,
ylabel={Coded BER},
title style={align=center},
axis x line*=bottom,
axis y line*=left,
legend columns=2,
legend style={at={(0.97,0.97)},anchor=north east,legend cell align=left,align=left,draw=black}
]
\addplot [color=color_em,very thick,mark=square,mark options={solid}]
  table[row sep=crcr]{%
  -9	0.178996682167053\\
  -7	0.0330078601837158\\
  -5	0.00370633602142334\\
  -3	0.000497221946716309\\
  -1	9.16719436645508e-05\\
  1	2.68220901489258e-05\\
  3	1.04506810506185e-05\\
};
\addlegendentry{EM-MC};

\addplot [color=color_em,very thick,dashed,mark=square,mark options={solid}]
  table[row sep=crcr]{%
-9	0.139207681020101\\
-7	0.0153633753458659\\
-5	0.000848929087320964\\
-3	6.63201014200846e-05\\
-1	1.06890996297201e-05\\
1	5.9207280476888e-06\\
3	3.05970509847005e-06\\
};
\addlegendentry{EM-SC};

\addplot [color=color_gamp,very thick,mark=triangle,mark options={solid,rotate=180}]
  table[row sep=crcr]{%
-9	0.17684809366862\\
-7	0.032036026318868\\
-5	0.00354635715484619\\
-3	0.000469247500101725\\
-1	8.69035720825195e-05\\
1	2.41994857788086e-05\\
3	1.02519989013672e-05\\
};
\addlegendentry{GAMP-MC};

\addplot [color=color_gamp,very thick,dashed,mark=triangle,mark options={solid,rotate=180}]
  table[row sep=crcr]{%
-9	0.138112147649129\\
-7	0.0149054924647013\\
-5	0.000797708829243978\\
-3	6.11146291097005e-05\\
-1	1.02122624715169e-05\\
1	6.07967376708984e-06\\
3	2.98023223876953e-06\\
};
\addlegendentry{GAMP-SC};

\addplot [color=color_bussgang,very thick,mark=o,mark options={solid}]
  table[row sep=crcr]{%
-9	0.184950709342957\\
-7	0.0365574757258097\\
-5	0.00456098715464274\\
-3	0.000668207804361979\\
-1	0.000135898590087891\\
1	4.57763671875e-05\\
3	2.24908192952474e-05\\
};
\addlegendentry{Buss.-MC};

\addplot [color=color_bussgang,very thick,dashed,mark=o,mark options={solid}]
  table[row sep=crcr]{%
-9	0.150791486104329\\
-7	0.0197595755259196\\
-5	0.00141473611195882\\
-3	0.000129620234171549\\
-1	2.24510828653971e-05\\
1	9.77516174316406e-06\\
3	7.66913096110026e-06\\
};
\addlegendentry{Buss.-SC};

\addplot [color=color_ignoring,very thick,mark=triangle,mark options={solid}]
  table[row sep=crcr]{%
-9	0.19683833916982\\
-7	0.0453597704569499\\
-5	0.0068051020304362\\
-3	0.00115338961283366\\
-1	0.000279347101847331\\
1	0.000103433926900228\\
3	5.48362731933594e-05\\
};
\addlegendentry{Ign.-MC};

\addplot [color=color_ignoring,very thick,dashed,mark=triangle,mark options={solid}]
  table[row sep=crcr]{%
-9	0.152105927467346\\
-7	0.0204607248306274\\
-5	0.00150938828786214\\
-3	0.000150163968404134\\
-1	2.98420588175456e-05\\
1	1.33514404296875e-05\\
3	9.29832458496094e-06\\
};
\addlegendentry{Ign.-SC};

\end{axis}
\end{tikzpicture}%
\caption{Coded BER of channel estimation techniques with EM equalization ($N_r\times N_t=10\times 2$, $N=32$ subcarriers, $T=4$ pilot blocks, CC rate $3/4$)}
\label{fig:paper_methods_chest_ber}
\end{figure}

Fig.~\ref{fig:paper_methods_chest_ber} shows the coded BER results for the same systems. Important conclusions can be drawn from these two figures. First, all methods saturate at a certain finite SNR, above which no further improvement in performance is obtained. This is due to the well-known dithering effect: some amount of noise is actually beneficial for quantized measurements. 

The iterative nonlinear methods (EM and GAMP) clearly outperform the linear techniques, and saturate at a better performance. In the OFDM case, this comes at the cost of computational complexity, as the linear estimators can perform equalization efficiently in a subcarrier-wise fashion (\ref{eq:bussgang_iid_mmse}).

The OFDM systems (solid curves) perform slightly worse than the single-carrier ones (dashed curves). The higher Peak-to-Average Power Ratio (PAPR) of the OFDM modulation makes the quantization noise more severe in this setting.

\subsection{Comparison of Algorithms for Data Equalization}
\begin{figure}[!t]\centering
%
%
\definecolor{color_em}{rgb}{0.00000,0.50000,0.00000}%
\definecolor{color_gamp}{rgb}{1.00000,0.00000,0.00000}%
\definecolor{color_bussgang}{rgb}{0.00000,0.00000,1.00000}%
\definecolor{color_ignoring}{rgb}{1.00000,0.500000,0.00000}%
\begin{tikzpicture}

\begin{axis}[%
width=0.972222\figurewidth,
height=\figureheight,
at={(0\figurewidth,0\figureheight)},
scale only axis,
every outer x axis line/.append style={black},
every x tick label/.append style={font=\color{black}},
xmin=-9,
xmax=3,
xtick={-9,-7,-5,-3,-1,1,3},
xlabel={SNR (dB)},
every outer y axis line/.append style={black},
every y tick label/.append style={font=\color{black}},
ymode=log,
ymin=4e-7,
ymax=0.7,
ytick={1e-6,1e-5,1e-4,1e-3,1e-2,1e-1},
yminorticks=true,
xmajorgrids,
ymajorgrids,
yminorgrids,
ylabel={Coded BER},
title style={align=center},
axis x line*=bottom,
axis y line*=left,
legend columns=2,
legend style={legend cell align=left,align=left,draw=black}
]
\addplot [color=color_em,very thick,mark=square,mark options={solid}]
  table[row sep=crcr]{%
-9	0.0684550205866496\\
-7	0.00741044680277507\\
-5	0.000671903292338053\\
-3	9.22679901123047e-05\\
-1	1.93913777669271e-05\\
1	4.92731730143229e-06\\
3	3.73522440592448e-06\\
};
\addlegendentry{EM-MC};

\addplot [color=color_em,very thick,dashed,mark=square,mark options={solid}]
  table[row sep=crcr]{%
-9	0.0479122002919515\\
-7	0.00272941589355469\\
-5	0.000121514002482096\\
-3	1.4185905456543e-05\\
-1	4.80810801188151e-06\\
1	1.98682149251302e-06\\
3	5.96046447753906e-07\\
};
\addlegendentry{EM-SC};

\addplot [color=color_gamp,very thick,mark=triangle,mark options={solid,rotate=180}]
  table[row sep=crcr]{%
-9	0.0680207808812459\\
-7	0.00728988647460938\\
-5	0.000644961992899577\\
-3	8.84135564168294e-05\\
-1	1.58945719401042e-05\\
1	4.76837158203125e-06\\
3	2.82128651936849e-06\\
};
\addlegendentry{GAMP-MC};

\addplot [color=color_gamp,very thick,dashed,mark=triangle,mark options={solid,rotate=180}]
  table[row sep=crcr]{%
-9	0.0482408205668131\\
-7	0.00277320543924967\\
-5	0.000119209289550781\\
-3	1.27951304117839e-05\\
-1	3.37759653727214e-06\\
1	1.62919362386068e-06\\
3	9.1393788655599e-07\\
};
\addlegendentry{GAMP-SC};

\addplot [color=color_bussgang,very thick,mark=o,mark options={solid}]
  table[row sep=crcr]{%
-9	0.0760863224665324\\
-7	0.0101664861043294\\
-5	0.00123647848765055\\
-3	0.000239332516988118\\
-1	6.63598378499349e-05\\
1	3.00010045369466e-05\\
3	1.87158584594727e-05\\
};
\addlegendentry{Buss.-MC};

\addplot [color=color_bussgang,very thick,dashed,mark=o,mark options={solid}]
  table[row sep=crcr]{%
-9	0.0515209039052327\\
-7	0.00330519676208496\\
-5	0.000172932942708333\\
-3	2.13781992594401e-05\\
-1	7.78834025065104e-06\\
1	4.52995300292969e-06\\
3	2.34444936116536e-06\\
};
\addlegendentry{Buss.-SC};

\addplot [color=color_ignoring,very thick,mark=triangle,mark options={solid}]
  table[row sep=crcr]{%
-9	0.0862418413162231\\
-7	0.0134026606877645\\
-5	0.00197625160217285\\
-3	0.00043634573618571\\
-1	0.000140547752380371\\
1	7.30355580647786e-05\\
3	4.57366307576497e-05\\
};
\addlegendentry{Ign.-MC};

\addplot [color=color_ignoring,very thick,dashed,mark=triangle,mark options={solid}]
  table[row sep=crcr]{%
-9	0.0584823290506999\\
-7	0.00464228789011637\\
-5	0.000317573547363281\\
-3	4.25577163696289e-05\\
-1	1.5099843343099e-05\\
1	7.31150309244792e-06\\
3	4.29153442382812e-06\\
};
\addlegendentry{Ign.-SC};

\end{axis}
\end{tikzpicture}%
\caption{Coded BER comparison of equalization techniques with perfect CSI ($N_r\times N_t=10\times 2$, $N=32$ subcarriers, CC rate $3/4$)}
\label{fig:paper_methods_eq}
\end{figure}
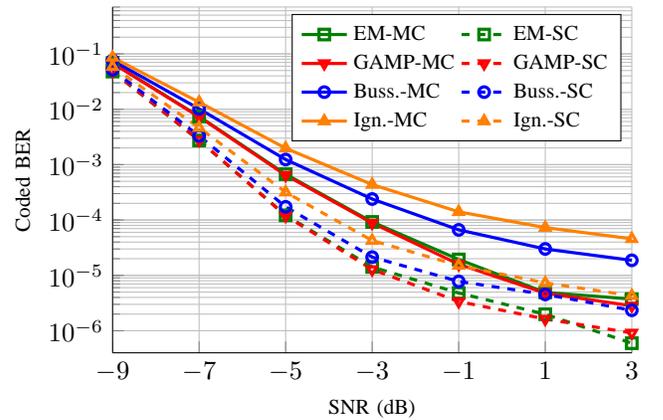

The same methods for OFDM and SC were compared for the equalization task in an $10\times 2$ system, assuming perfect CSI. The results are plotted in Fig.~\ref{fig:paper_methods_eq}. Again, SC beats MC, and the nonlinear methods perform better than the linear ones.

\subsection{Number of Pilots}
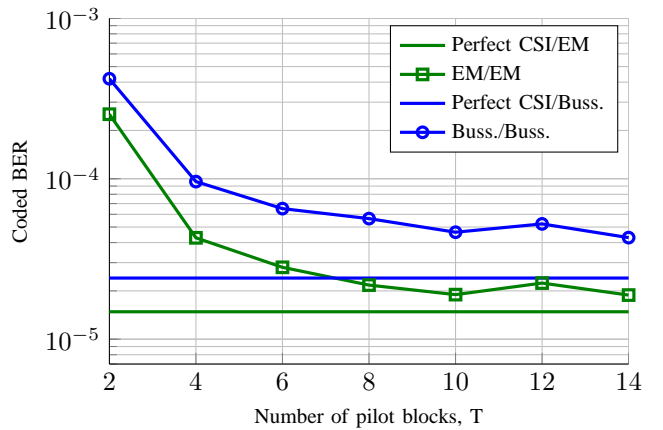
\begin{figure}[!t]\centering
%
%
\definecolor{color_em}{rgb}{0.00000,0.50000,0.00000}%
\definecolor{color_gamp}{rgb}{1.00000,0.00000,0.00000}%
\definecolor{color_bussgang}{rgb}{0.00000,0.00000,1.00000}%
\begin{tikzpicture}

\begin{axis}[%
width=0.972222\figurewidth,
height=\figureheight,
at={(0\figurewidth,0\figureheight)},
scale only axis,
every outer x axis line/.append style={black},
every x tick label/.append style={font=\color{black}},
xmin=2,
xmax=14,
xtick={2,4,6,8,10,12,14},
xlabel={Number of pilot blocks, T},
every outer y axis line/.append style={black},
every y tick label/.append style={font=\color{black}},
ymode=log,
ymin=7e-6,
ymax=1e-3,
yminorticks=true,
xmajorgrids,
ymajorgrids,
yminorgrids,
ylabel={Coded BER},
title style={align=center},
axis x line*=bottom,
axis y line*=left,
legend style={legend cell align=left,align=left,draw=black}
]
\addplot [color=color_em,very thick]
  table[row sep=crcr]{%
  2	1.48216883341471e-05\\
  14	1.48216883341471e-05\\
};
\addlegendentry{Perfect CSI/EM};

\addplot [color=color_em,very thick,mark=square,mark options={solid}]
  table[row sep=crcr]{%
  2	0.000252445538838704\\
  4	4.28358713785807e-05\\
  6	2.80539194742839e-05\\
  8	2.17358271280924e-05\\
  10	1.89542770385742e-05\\
  12	2.23716100056966e-05\\
  14	1.88350677490234e-05\\
};
\addlegendentry{EM/EM};



\addplot [color=color_bussgang,very thick]
  table[row sep=crcr]{%
  2	2.40405400594076e-05\\
  14	2.40405400594076e-05\\
};
\addlegendentry{Perfect CSI/Buss.};

\addplot [color=color_bussgang,very thick,mark=o,mark options={solid}]
  table[row sep=crcr]{%
  2	0.000420014063517253\\
  4	9.60032145182292e-05\\
  6	6.51280085245768e-05\\
  8	5.64257303873698e-05\\
  10	4.64121500651042e-05\\
  12	5.22931416829427e-05\\
  14	4.29550806681315e-05\\
};
\addlegendentry{Buss./Buss.};

\end{axis}
\end{tikzpicture}%
\caption{Number of pilot blocks required for SC estimation and equalization ($N_r\times N_t=10\times 2$, $N=32$ subcarriers, $\mathrm{SNR}=-3\dB$, CC rate $3/4$)}
\label{fig:paper_pilots_ber}
\end{figure}

In the third experiment, the SNR was fixed at $-3\dB$, and the EM and Bussgang methods were compared in terms of BER vs number of pilot blocks. In the two cases, both equalization and channel estimation were performed with the corresponding method in an SC system. The curves were compared with the perfect CSI case. The results in Fig.~\ref{fig:paper_pilots_ber} show that, the Bussgang saturates at a worse performance than EM. Additionally, it is seen that $4$ pilot blocks are enough to perform reasonably close to saturation.

\subsection{Higher Order Modulation}
\begin{figure}[!t]\centering
%
%
\definecolor{color_em}{rgb}{0.00000,0.50000,0.00000}%
\definecolor{color_bussgang}{rgb}{0.00000,0.00000,1.00000}%
\begin{tikzpicture}

\begin{axis}[%
width=0.95092\figurewidth,
height=\figureheight,
at={(0\figurewidth,0\figureheight)},
scale only axis,
every outer x axis line/.append style={black},
every x tick label/.append style={font=\color{black}},
xmin=-15,
xmax=15,
xtick={-15,-10,-5,0,5,10,15},
xlabel={SNR (dB)},
every outer y axis line/.append style={black},
every y tick label/.append style={font=\color{black}},
ymode=log,
ymin=8e-8,
ymax=0.5,
ytick={1e-7,1e-6,1e-5,1e-4,1e-3,1e-2,1e-1},
yminorticks=true,
xmajorgrids,
ymajorgrids,
yminorgrids,
ylabel={Coded BER},
title style={align=center},
axis x line*=bottom,
axis y line*=left,
legend style={at={(0.97,0.97)},anchor=north east,legend cell align=left,align=left,draw=black}
]
\addplot [color=color_em, very thick,solid,mark=o,mark options={solid}]
  table[row sep=crcr]{%
-15	0.477511485417684\\
-10	0.0463194052378337\\
-5	3.814697265625e-05\\
0	2.88751390245226e-06\\
5	1.11262003580729e-06\\
10	9.27183363172743e-07\\
15	3.97364298502604e-07\\
};
\addlegendentry{EM ($8$-PSK)};

\addplot [color=color_em,dashed,very thick,mark=square,mark options={solid}]
  table[row sep=crcr]{%
-15	0.484538290235731\\
-10	0.0845048162672255\\
-5	0.000113301806979709\\
0	4.39749823676215e-06\\
5	1.80138481987847e-06\\
10	1.9073486328125e-06\\
15	2.70207722981771e-06\\
};
\addlegendentry{EM ($8$-QAM)};

\addplot [color=color_bussgang,very thick,mark=o,mark options={solid}]
  table[row sep=crcr]{%
-15	0.478221654891968\\
-10	0.0618696212768555\\
-5	0.000159660975138346\\
0	1.30865308973524e-05\\
5	8.10623168945312e-06\\
10	5.61608208550347e-06\\
15	6.30484686957465e-06\\
};
\addlegendentry{Buss. ($8$-PSK)};

\addplot [color=color_bussgang,dashed,very thick,mark=square,mark options={solid}]
  table[row sep=crcr]{%
-15	0.492243872748481\\
-10	0.161263995700412\\
-5	0.00167748663160536\\
0	0.000162283579508464\\
5	9.31686825222439e-05\\
10	7.35123952229818e-05\\
15	8.52743784586589e-05\\
};
\addlegendentry{Buss. ($8$-QAM)};

\end{axis}
\end{tikzpicture}%
\caption{Coded BER comparison of modulation schemes. The legend entries give the method for both channel estimation and equalization ($N_r\times N_t=24\times 2$, $N=32$ subcarriers, $T=4$ pilot blocks, CC rate=$3/4$)}
\label{fig:paper_mod_ber}
\end{figure}
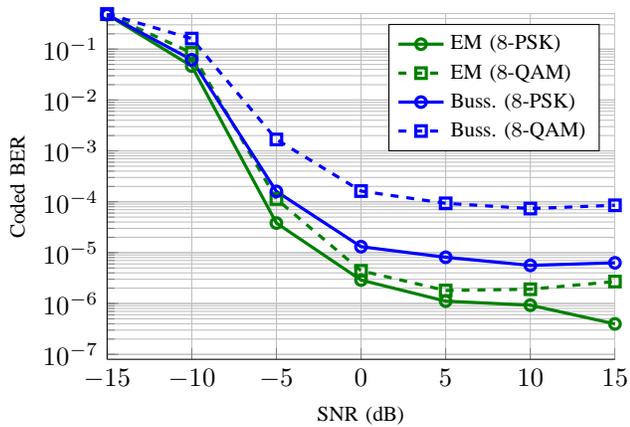

The fourth experiment aimed at assessing the viability of using higher order modulations at the transmitter, while keeping 1-bit quantization at the receiver. For this purpose, an SC system with more receive antennas ($24\times 2$) was simulated, and the BER with $8$-QAM and $8$-PSK was compared with the EM and the Bussgang estimators (the estimators are used for both channel estimation and equalization). The results are shown in Fig.~\ref{fig:paper_mod_ber}. If $3$ bits per symbol are required, $8$-PSK is more effective for the 1-bit quantized systems than $8$-QAM. This is because $1$-bit quantization causes more damage to the signal amplitude that to its phase. For low modulation orders, therefore, it is better to use PSK modulations that only store information in the phase. 

\subsection{Full System with 3GPP Channel Model}
\begin{figure}[!t]\centering
%
%
\definecolor{color_em}{rgb}{0.00000,0.50000,0.00000}%
\definecolor{color_gamp}{rgb}{1.00000,0.00000,0.00000}%
\definecolor{color_bussgang}{rgb}{0.00000,0.00000,1.00000}%
\definecolor{color_ignoring}{rgb}{1.00000,0.500000,0.00000}%
\begin{tikzpicture}

\begin{axis}[%
width=0.95092\figurewidth,
height=\figureheight,
at={(0\figurewidth,0\figureheight)},
scale only axis,
every outer x axis line/.append style={black},
every x tick label/.append style={font=\color{black}},
xmin=-9,
xmax=3,
xtick={-9,-7,-5,-3,-1,1,3},
xlabel={SNR (dB)},
every outer y axis line/.append style={black},
every y tick label/.append style={font=\color{black}},
ymode=log,
ymin=1e-6,
ymax=1,
ytick={1e-6,1e-5,1e-4,1e-3,1e-2,1e-1,1},
yminorticks=true,
xmajorgrids,
ymajorgrids,
yminorgrids,
ylabel={Coded BER},
title style={align=center},
axis x line*=bottom,
axis y line*=left,
legend columns=2,
legend style={legend cell align=left,align=left,draw=black}
]
\addplot [color=color_em,very thick,mark=square,mark options={solid}]
  table[row sep=crcr]{%
-9	0.0195126533508301\\
-7	0.00252405802408854\\
-5	0.000413417816162109\\
-3	8.88506571451823e-05\\
-1	1.9232432047526e-05\\
1	1.08083089192708e-05\\
3	7.62939453125e-06\\
};
\addlegendentry{EM-MC};

\addplot [color=color_em, very thick, dashed,mark=square,mark options={solid}]
  table[row sep=crcr]{%
-9	0.00574096043904622\\
-7	0.000393867492675781\\
-5	3.57627868652344e-05\\
-3	1.17619832356771e-05\\
-1	1.9073486328125e-06\\
1	1.43051147460938e-06\\
3	1.74840291341146e-06\\
};
\addlegendentry{EM-SC};

\addplot [color=color_gamp,very thick,mark=triangle,mark options={solid,rotate=180}]
  table[row sep=crcr]{%
-9	0.0197796821594238\\
-7	0.00258111953735352\\
-5	0.000409762064615885\\
-3	9.5367431640625e-05\\
-1	1.5099843343099e-05\\
1	1.00135803222656e-05\\
3	7.94728597005208e-06\\
};
\addlegendentry{GAMP-MC};

\addplot [color=color_gamp, very thick, dashed,mark=triangle,mark options={solid,rotate=180}]
  table[row sep=crcr]{%
-9	0.00616852442423503\\
-7	0.000439564387003581\\
-5	4.70081965128581e-05\\
-3	1.23182932535807e-05\\
-1	5.7220458984375e-06\\
1	4.76837158203125e-06\\
3	3.814697265625e-06\\
};
\addlegendentry{GAMP-SC};

\addplot [color=color_bussgang,very thick,mark=o,mark options={solid}]
  table[row sep=crcr]{%
-9	0.0314534505208333\\
-7	0.0058892567952474\\
-5	0.00144449869791667\\
-3	0.000428040822347005\\
-1	0.000195980072021484\\
1	0.000123500823974609\\
3	9.63211059570312e-05\\
};
\addlegendentry{Buss.-MC};

\addplot [color=color_bussgang, very thick, dashed,mark=o,mark options={solid}]
  table[row sep=crcr]{%
-9	0.00959030787150065\\
-7	0.000880877176920573\\
-5	0.000135580698649089\\
-3	5.27699788411458e-05\\
-1	1.79608662923177e-05\\
1	1.23977661132812e-05\\
3	4.76837158203125e-06\\
};
\addlegendentry{Buss.-SC};

\addplot [color=color_ignoring, very thick,mark=triangle,mark options={solid}]
  table[row sep=crcr]{%
-9	0.0496490796407064\\
-7	0.013237476348877\\
-5	0.0044865608215332\\
-3	0.00190528233846029\\
-1	0.00120639801025391\\
1	0.000902970631917318\\
3	0.000733693440755208\\
};
\addlegendentry{Ign.-MC};

\addplot [color=color_ignoring, very thick, dashed,mark=triangle,mark options={solid}]
  table[row sep=crcr]{%
-9	0.0105512936909993\\
-7	0.00106922785441081\\
-5	0.000172615051269531\\
-3	7.72476196289062e-05\\
-1	2.41597493489583e-05\\
1	2.43186950683594e-05\\
3	1.5099843343099e-05\\
};
\addlegendentry{Ign.-SC};

\end{axis}
\end{tikzpicture}%
\caption{Coded BER comparison with 3GPP channel model. The legend entries give the method for both channel estimation and equalization ($N_r\times N_t=10\times 2$, $N=32$ subcarriers, $T=4$ pilot blocks, CC rate=$3/4$)}
\label{fig:paper_3gpp_ber}
\end{figure}
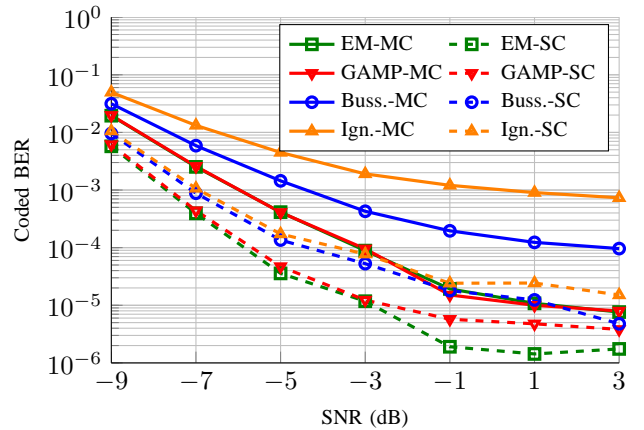

The estimation methods discussed in this paper all assume Gaussian i.i.d. channel taps. In the last experiment, the algorithms were tested using a more realistic channel model: the Extended Pedestrian A model from 3GPP TS 36.101~\cite{3gpp_model}. The results for the four methods with both OFDM and SC are given in Fig.~\ref{fig:paper_3gpp_ber}, where again each algorithm is used both for channel estimation and equalization. It is seen that the GAMP algorithm suffers slightly more degradation in the SC case. This is because it relies on the assumption that the transform matrix $\mathbf{A}$ has i.i.d. Gaussian entries (see Section III.A of~\cite{rangan_GAMP}). The channel models for the OFDM case (\ref{eq:chest_sensing}) and (\ref{eq:eq_sensing}) are closer to this assumption than the convolution matrices for the single-carrier model. For OFDM, however, it gives the best performance at a lower complexity than EM.

The other methods are also seen to degrade with respect to the results with i.i.d. Gaussian channels (Figs.~\ref{fig:paper_methods_chest_nmse}-\ref{fig:paper_pilots_ber}), but they turn out to be much more robust than GAMP.

\section{Conclusions}\label{sec:conclusions}
An analysis and comparison of different channel estimation and data equalization techniques for a frequency-selective MIMO system with 1-bit quantization at the receiver was carried out. Channel models were derived for both single-carrier and multi-carrier schemes. A Cram\'er-Rao bound for the estimator variance was obtained. Two existing nonlinear iterative algorithms were adapted to the estimation problem. A linear estimator based on the Bussgang theorem was proposed, which greatly reduces complexity in the multi-carrier case and outperforms the quantization-ignoring linear estimator.

Through simulations, it was shown that all algorithms have a performance peak at a finite SNR value. The nonlinear methods have an unavoidable advantage over the linear ones at high SNR. Single Carrier performs better than OFDM, but an OFDM system with Bussgang estimation is the best solution for medium SNR if computational complexity is an issue.

It was also seen that $8$-PSK outperforms $8$-QAM if 3 bits per symbol are required. This still requires an increase in the number of receive antennas.

There are still numerous open challenges for frequency-selective MIMO channels. The application of joint channel and data estimation (JCD)~\cite{wen_jcd} to the frequency-selective case, the derivation of a Cram\'er-Rao bound for this problem, the design of pilot sequences, and the adaptation of the algorithms to correlated channel models, are left for future work. 


%
\appendices
\section{Derivation of the Fisher Information Matrix (\ref{eq:cr_fisher_final})}\label{sec:fisher}

Let us denote the dimensions of $\tilde{\mathbf{A}}$ in (\ref{eq:cr_real_valued}) by $K$ and $P$, so that $\tilde{\mathbf{A}}\inC{K\times P}$. For the case in which $\mathcal{Q}\left(\cdot\right)$ applies 1-bit quantization (\ref{eq:Q}) and $\mathbf{w}$ has uncorrelated Gaussian samples with variances $\sigma_k^2, k\inset{1}{K}$, we have:
\begin{equation}
p_{\tilde{\mathbf{y}}\given{\tilde{\mathbf{h}}}}\left(\tilde{\mathbf{y}}\given{\tilde{\mathbf{h}}}\right)=\prod_{k=1}^K\mathrm{\Phi}\left(\frac{\tilde{y}_k\sum\limits_{p=1}^P \tilde{a}_{kp} \tilde{h}_p}{\sigma_k}\right)=\prod_{k=1}^K\mathrm{\Phi}\left(\eta_k\right),
\end{equation}
where $\mathrm{\Phi}\left(x\right)\triangleq\int_{-\infty}^{x}\frac{1}{\sqrt{2\pi}}e^{-v^2/2}\diff{v}$ denotes the standard cumulative Gaussian distribution function, and we define $\eta_k$ as the content of the parenthesis to simplify notation. The Fisher information matrix then yields:
\begin{align}
&\tilde{\mathbf{I}}\left(\tilde{\mathbf{h}}\right)=\mathrm{E}\left\{-\frac{\partial^2}{\partial\tilde{\mathbf{h}}^T\partial\tilde{\mathbf{h}}}\ln p_{\tilde{\mathbf{y}}\given{\tilde{\mathbf{h}}}}\left(\tilde{\mathbf{y}}\given{\tilde{\mathbf{h}}}\right)\right\}\nonumber\\
&=\mathrm{E}\left\{\tilde{\mathbf{A}}^T\mathrm{diag}\left\{\frac{1}{\sigma_k^2}\left(\frac{\eta_k\phi\left(\eta_k\right)}{\mathrm{\Phi}\left(\eta_k\right)}+\left(\frac{\phi\left(\eta_k\right)}{\mathrm{\Phi}\left(\eta_k\right)}\right)^2\right)\right\}_{k=1}^K\tilde{\mathbf{A}}\right\},
\label{eq:cr_fisher}
\end{align}
where $\phi\left(x\right)\triangleq\frac{1}{\sqrt{2\pi}}e^{-x^2/2}$ is the standard Gaussian density function. Now, we take the expectation over $\tilde{\mathbf{y}}\given{\tilde{\mathbf{h}}}$:
\begin{multline}
\tilde{\mathbf{I}}(\tilde{\mathbf{h}})=\tilde{\mathbf{A}}^T\mathrm{diag}\left\{\frac{1}{\sigma_k^2}\sum_{\tilde{\mathbf{y}}\in\left\{-1,1\right\}^K}^{}\left[\left(\frac{\eta_k\phi\left(\eta_k\right)}{\mathrm{\Phi}\left(\eta_k\right)}+\right.\right.\right.\\\left.\left.\left.+\left(\frac{\phi\left(\eta_k\right)}{\mathrm{\Phi}\left(\eta_k\right)}\right)^2\right)\prod_{k'=1}^{K}\mathrm{\Phi}\left(\eta_{k'}\right)\right]\right\}_{k=1}^K\tilde{\mathbf{A}}.
\label{eq:cr_fisher_pre}
\end{multline}
Note that $\left(\frac{\eta_k\phi\left(\eta_k\right)}{\mathrm{\Phi}\left(\eta_k\right)}+\left(\frac{\phi\left(\eta_k\right)}{\mathrm{\Phi}\left(\eta_k\right)}\right)^2\right)$ depends on $\tilde{y}_k$ (through $\eta_k$), but not on any other component of $\tilde{\mathbf{y}}$. After reorganizing the sum of products as a product of sums, this implies that all the sums corresponding to the other components equal $1$, and (\ref{eq:cr_fisher_pre}) reduces to (\ref{eq:cr_fisher_final}).

\section{Derivation of the Expectation Step in EM (\ref{eq:em_e})}\label{sec:em}
In the following, we drop the iteration index $i$ for clarity. Using Bayes rule, we obtain the distribution of $\mathbf{z}\given{\mathbf{y},\hat{\mathbf{h}}}$:
\begin{align}
p_{\mathbf{z}\given{\mathbf{y},\hat{\mathbf{h}}}}\left(\mathbf{z}\given{\mathbf{y},\hat{\mathbf{h}}}\right)&=\frac{p_{\mathbf{z}\given{\hat{\mathbf{h}}}}(\mathbf{z}\given{\hat{\mathbf{h}}})p_{\mathbf{y}\given{\mathbf{z},\hat{\mathbf{h}}}}(\mathbf{y}\given{\mathbf{z},\hat{\mathbf{h}}})}{p_{\mathbf{y}\given{\hat{\mathbf{h}}}}(\mathbf{y}\given{\hat{\mathbf{h}}})}\nonumber\\
&=\frac{p_{\mathbf{w}}\left(\mathbf{z}-\mathbf{A}\hat{\mathbf{h}}\right)\mathbf{1}\left\{\mathbf{z}\in\mathcal{Q}^{-1}\left(\mathbf{y}\right)\right\}}{\int\limits_{\mathbf{w}\in\mathcal{Q}^{-1}\left(\mathbf{y}\right)-\mathbf{A}\hat{\mathbf{h}}}p_\mathbf{w}(\mathbf{w})\mathrm{d}\mathbf{w}},
\label{eq:em_bayes}
\end{align}
where $\mathbf{1}\left\{s\right\}$ is an indicator function with value $1$ if $s$ is true, and $0$ otherwise. The set $\mathcal{Q}^{-1}\left(\mathbf{y}\right)$ is defined as:
\begin{equation}
\mathcal{Q}^{-1}\left(\mathbf{y}\right)\triangleq \left\{\mathbf{z}\inC{K}:\mathcal{Q}\left(\mathbf{z}\right)=\mathbf{y}\right\},
\end{equation}
and $\mathcal{Q}^{-1}\left(\mathbf{y}\right)-\mathbf{A}\hat{\mathbf{h}}$ is the translation of $\mathcal{Q}^{-1}\left(\mathbf{y}\right)$ by $-\mathbf{A}\hat{\mathbf{h}}$.

Now, we apply the expectation operator to (\ref{eq:em_bayes}):
\begin{align}
&\mathrm{E}\left(\mathbf{z}\given{\mathbf{y},\hat{\mathbf{h}}}\right)=\frac{\int\limits_{\mathbf{z}\in\mathcal{Q}^{-1}\left(\mathbf{y}\right)}\mathbf{z}\;p_\mathbf{w}\left(\mathbf{z}-\mathbf{A}\hat{\mathbf{h}}\right)\mathrm{d}\mathbf{z}}{\int\limits_{\mathbf{w}\in\mathcal{Q}^{-1}\left(\mathbf{y}\right)-\mathbf{A}\hat{\mathbf{h}}}p_\mathbf{w}(\mathbf{w})\mathrm{d}\mathbf{w}}\nonumber\\
&=\mathbf{A}\hat{\mathbf{h}}+\frac{\int\limits_{\mathbf{w}\in\mathcal{Q}^{-1}\left(\mathbf{y}\right)-\mathbf{A}\hat{\mathbf{h}}}\mathbf{}\;\mathbf{w}p_\mathbf{w}\left(\mathbf{w}\right)\mathrm{d}\mathbf{w}}{\int\limits_{\mathbf{w}\in\mathcal{Q}^{-1}\left(\mathbf{y}\right)-\mathbf{A}\hat{\mathbf{h}}}p_\mathbf{w}(\mathbf{w})\mathrm{d}\mathbf{w}}=\mathbf{A}\hat{\mathbf{h}}+\hat{\mathbf{w}}.
\label{eq:E_z_yh_generic}
\end{align}
Assuming that the noise $\mathbf{w}$ is Gaussian and uncorrelated with variances $\sigma_k^2, k\inset{1}{K}$, we have:
\begin{equation}
p_{\mathbf{w}}\left(\mathbf{w}\right)=\prod\limits_{k=1}^K \left(\frac{1}{\sigma_k\sqrt{\pi}}e^{-\frac{\Re\left\{w_k\right\}^2}{\sigma_k^2}}\frac{1}{\sigma_k\sqrt{\pi}}e^{-\frac{\Im\left\{w_k\right\}^2}{\sigma_k^2}}\right),
\label{eq:em_p_w}
\end{equation}
where $w_k$ denotes the $k$-th element of $\mathbf{w}$.

The integral in the numerator of (\ref{eq:E_z_yh_generic}) is vector-valued. Note that $p_{\mathbf{w}}\left(\mathbf{w}\right)$ is separable. Therefore, for the $k$-th component of the numerator, all dimensions will cancel out except for the $k$-th one, yielding (\ref{eq:em_e}).

\section{Derivation of the Nonlinear Steps of GAMP-MMSE}\label{app:gamp_steps}
The nonlinear steps of GAMP are elementwise independent, and therefore we will derive them for an individual sample $x_i$ (input) or $y_j$ (output). We drop the sample index for clarity.

\subsection{Gaussian Input Step}
For channel estimation, the input is assumed to be Gaussian uncorrelated with variance $\sigma_x^2$ (possibly different for each sample). The inner variable $r$ of GAMP is defined as a noisy estimate of the input variable $x$, with Gaussian-distributed uncorrelated noise $v$~\cite{rangan_GAMP}:
\begin{equation}
r=x+v,\qquad \mathrm{with}\  v\sim\mathcal{N}_{\mathbb{C}}\left(0, \tau^ r\right).
\label{eq:r}
\end{equation}
The input nonlinear function $g_{in}$ is then given by:
\begin{equation}
g_{in}\left(i, r, \tau^r\right)=\mathrm{E}\left\{x\given{r}\right\}.
\label{eq:g_in_def}
\end{equation}
In our problem, we have $x\sim\mathcal{N}_{\mathbb{C}}\left(0,\sigma_x^2\right)$. Using Bayes' Rule, we obtain:
\begin{align}
p_{x\given{r}}\left(x\given{ r}\right)&=\frac{p_{x}\left(x\right)p_{r\given{x}}\left(r\given{x}\right)}{p_{r}\left(r\right)}\label{eq:px_qr}\\ &=\frac{\frac{1}{\sigma_x^2\pi}e^{-\frac{\left|x\right|^2}{\sigma_x^2}}\frac{1}{\pi\tau^r}e^{-\frac{\left|r-x\right|^2}{\tau^r}}}{\frac{1}{\pi\left(\sigma_x^2+\tau^r\right)}e^{-\frac{\left|r\right|^2}{\left(\sigma_x^2+\tau^r\right)}}}\nonumber\\ &=\frac{1}{\pi\sigma_{x\given{r}}^2}e^{-\frac{\left|x-\mu_{x\given{r}}\right|^2}{\sigma_{x\given{r}}^2}},
\label{eq:px_r}
\end{align}
where 
\begin{equation}
\mu_{x\given{r}}=\frac{\sigma_x^2r}{\sigma_x^2+\tau^r},
\end{equation}
\begin{equation}
\sigma_{x\given{r}}^2=\frac{\sigma_x^2\tau_r}{\sigma_x^2+\tau^r}.
\end{equation}
The PDF in (\ref{eq:px_r}) is Gaussian with mean $\mu_{x\given{r}}$ and variance $\sigma_{x\given{r}}^2$. From (\ref{eq:g_in_def}), $g_{in}$ is equal to $\mu_{x\given{r}}$, which results in the input nonlinear functions (\ref{eq:gamp_g_in}) and (\ref{eq:gamp_dg_in}).

\subsection{Constellation Input Step}
In the equalization problems, the constellation of the input $x$ is known. Let us denote the constellation points by $\overline{x}_a, a\inset{1}{A}$, where $A$ is the constellation order. The probability of $\overline{x}_a$ is denoted by $P_a$. By applying (\ref{eq:r}) and (\ref{eq:px_qr}), we obtain:
\begin{equation}
p_{x\given{r}}\left(\overline{x}_a\given{r}\right)=\frac{P_a\frac{1}{\tau^r\pi}e^{-\frac{\left|r-\overline{x}_a\right|^2}{\tau^r}}}{\sum\limits_{a=1}^{A}P_a\frac{1}{\tau^r\pi}e^{-\frac{\left|r-\overline{x}_a\right|^2}{\tau^r}}},
\end{equation}
The expectation $\mathrm{E}\left\{x\given{r}\right\}$ is computed by averaging over $x$, yielding (\ref{eq:gamp_g_in_c}) and (\ref{eq:gamp_dg_in_c}).

\subsection{Quantized Output Step}
Let us turn now to the output steps. In this case, the relevant inner variable is $p$, which is defined such that:
\begin{equation}
z=p+u,\qquad \mathrm{with}\ u\sim\mathcal{N}_{\mathbb{C}}\left(0,\tau^p\right),
\end{equation}
where $u$ is independent from $p$. The output nonlinear function $g_{out}$ is then defined as:
\begin{equation}
g_{out}\left(i,p,y,\tau^p\right)=\frac{2}{\tau^p}\left(\mathrm{E}\left\{z\given{p,y}\right\}-p\right)=\frac{2}{\tau^p}\mathrm{E}\left\{u\given{p,y}\right\}.
\label{eq:g_out_def}
\end{equation}

In our 1-bit quantized case, the real and imaginary parts of the problem are independent. Therefore, the expectation can be taken separately for the two components, and we only derive the result for the real part:
\begin{equation}
y=\mathcal{Q}\left(z+w\right)=\mathcal{Q}\left(p+u+w\right),
\end{equation}
where $u\sim\mathcal{N}\left(0, \tau^p/2\right)$. The real-valued noise $w\sim\mathcal{N}\left(0, \sigma_w^2/2\right)$ is uncorrelated with $p$ and $u$. Now, we can use Bayes' rule again to obtain the joint PDF of $u,w\given{p,y}$:
\begin{multline}
p_{u,w\given{p,y}}\left(u,w\given{p,y}\right)=\frac{p_{u,w\given{p}}\left(u,w\given{p}\right)p_{y\given{u,w,p}}\left(y\given{u,w,p}\right)}{p_{y\given{p}}\left(y\given{p}\right)}\\
=\frac{\frac{1}{\pi\sigma_w\sqrt{\tau^p}}e^{-\frac{u^2}{\tau^p}-\frac{w^2}{\sigma_w^2}}\mathbf{1}\left\{\left(p+u+w\right)y\ge 0\right\}}{\mathrm{\Phi}\left(\frac{yp\sqrt{2}}{\sqrt{\tau^p+\sigma_w^2}}\right)}.
\label{eq:pvw_py}
\end{multline}
Now, we marginalize over $w$ and average over $u$:
\begin{equation}
\mathrm{E}\left\{u\given{p,y}\right\}=\frac{1}{\mathrm{\Phi}\left(\eta\right)}\int_{-\infty}^{\infty}\int_{-\infty}^{\infty}up_{u,w\given{p,y}}\left(u,w\given{p,y}\right)\diff{u}\diff{w}.
\label{eq:E_v_py}
\end{equation}
where $\eta=yp\sqrt{2/\left(\tau^p+\sigma_w^2\right)}$. By appropriately expressing the indicator function, (\ref{eq:E_v_py}) can be written as:
\begin{multline}
\mathrm{E}\left\{u\given{p,y}\right\} \\ =\frac{1}{\mathrm{\Phi}\left(\eta\right)}\int_{-\infty}^{\infty}\frac{1}{\sigma_w\sqrt{\pi}}e^{-\frac{w^2}{\sigma_w^2}}y\int_{-p-w}^{\infty}\frac{u}{\sqrt{\pi\tau^p}}e^{-\frac{u^2}{\tau^p}}\diff{u}\diff{w}.
\label{eq:Ev_py}
\end{multline}
The integral along $u$ has the limits corresponding to $y=1$ (see the indicator function $\mathbf{1}\left\{\ldots\right\}$ in (\ref{eq:pvw_py})). For the case $y=-1$, we have used the property that the integrand $f(u)$ is an odd function, and therefore $\int_{-p-w}^{\infty}f(u)\diff{u}=-\int_{-\infty}^{-p-w}f(u)\diff{u}$, which accounts for the pre-multiplying term $y$. The solution to (\ref{eq:Ev_py}) is:
\begin{align}
\mathrm{E}\left\{u\given{p,y}\right\}&=\frac{1}{\mathrm{\Phi}\left(\eta\right)}\int_{-\infty}^{\infty}\frac{1}{\sigma_w\sqrt{\pi}}e^{-\frac{w^2}{\sigma_w^2}}y\frac{\sqrt{\tau^p}}{\sqrt{\pi}}e^{-\frac{\left(p+w\right)^2}{\tau^p}}\diff{w} \nonumber \\
&=\frac{y\tau^p}{\sqrt{2\left(\sigma_w^2+\tau^p\right)}}\frac{\phi\left(\eta\right)}{\mathrm{\Phi}\left(\eta\right)},
\end{align}
which, plugged into ($\ref{eq:g_out_def}$), gives the output nonlinear step functions (\ref{eq:gamp_g_out}) and (\ref{eq:gamp_dg_out}).

%
%

\ifCLASSOPTIONcaptionsoff
  \newpage
\fi



\bibliographystyle{IEEEtran}
\bibliography{paper_arxiv}
\end{document}